\documentclass[pdflatex,sn-mathphys-num]{sn-jnl}

\usepackage{graphicx}
\usepackage{amsmath,amssymb}
\usepackage{bm}
\usepackage[title]{appendix}
\usepackage{textcomp}
\usepackage{booktabs}
\usepackage{array}
\usepackage{microtype}
\numberwithin{equation}{section}
\newcommand{\Tf}{T^{(\mathrm f)}}
\newcommand{\Tem}{T^{(\mathrm{em})}}
\newcommand{\Bf}{\mathcal{B}}
\newcommand{\kmsMpc}{\ensuremath{\mathrm{km\,s^{-1}\,Mpc^{-1}}}}

\begin{document}
\raggedbottom

\title[Magnetized strange quark matter in hypersurface-homogeneous $f(R,T)$ cosmology]{Magnetized strange quark matter in hypersurface-homogeneous $f(R,T)$ cosmology: physical closure, isotropization and late-time observational constraints}

\author[1]{\fnm{D. P.} \sur{Tadas}}\email{dtadas144@rediffmail.com}

\author*[2]{\fnm{S. V.} \sur{Gore}}\email{swapnilgore411@gmail.com}

\author[3]{\fnm{S. P.} \sur{Hatkar}}\email{schnhatkar@gmail.com}

\author[4]{\fnm{S. D.} \sur{Katore}}\email{katore777@gmail.com}

\affil[1]{\orgdiv{Department of Mathematics}, \orgname{Toshniwal Arts, Commerce and Science College}, \orgaddress{\city{Sengaon}, \postcode{431542}, \state{Maharashtra}, \country{India}}}

\affil*[2]{\orgdiv{Department of Mathematics}, \orgname{Indira Gandhi Kala Mahavidyalaya}, \orgaddress{\city{Ralegaon}, \postcode{445402}, \state{Maharashtra}, \country{India}}}

\affil[3]{\orgdiv{Department of Mathematics}, \orgname{A.E.S. Arts, Commerce and Science College}, \orgaddress{\city{Hingoli}, \postcode{431513}, \state{Maharashtra}, \country{India}}}

\affil[4]{\orgdiv{Department of Mathematics}, \orgname{Sant Gadge Baba Amravati University}, \orgaddress{\city{Amravati}, \postcode{444602}, \state{Maharashtra}, \country{India}}}

\abstract{Cosmological models of magnetized strange quark matter in $f(R,T)$ gravity have so far relied on kinematic ansatzes and lacked observational tests. We revisit them for $f(R,T)=R+2\mu T$ on a hypersurface-homogeneous background containing the locally rotationally symmetric Bianchi I, Kantowski--Sachs and Bianchi III geometries. Treated exactly, the Maxwell field couples minimally. Because kinematic ansatzes contradict a magnetized source, we close the field equations with the bag-model equation of state and magnetic flux freezing and, adding dust and radiation, solve them exactly in limiting cases and numerically in general. Without primordial shear, the field-induced anisotropy keeps, as in general relativity, a nearly constant radiation-era level for $\mu\ge0$, then decays as the model approaches de Sitter space; an assumed present field of 1 nG leaves $(\sigma/H)_0\simeq5.8\times10^{-11}$ today. A joint fit to 32 cosmic chronometers, 1590 Pantheon+ supernova light curves and baryon acoustic oscillations from the second data release of the Dark Energy Spectroscopic Instrument gives $\mu=1.12^{+0.65}_{-0.58}$ and $H_0=69.1\pm3.2$~\kmsMpc, and the null, weak and dominant energy conditions hold throughout. Relative to the flat $\Lambda$ cold dark matter model, the Akaike (AIC) and Bayesian (BIC) information criteria give $\Delta\mathrm{AIC}=-1.4$ and $\Delta\mathrm{BIC}=+4.0$, and the Bayes factor is inconclusive. The fitted bag term, $B_c^{1/4}=2.15$ meV, is an effective vacuum energy, not the quantum chromodynamics bag constant. The preference for $\mu>0$ depends on the supernova sample and the prior, a constant coupling would shift matter--radiation equality, and the sign of $\mu$ depends on the matter Lagrangian; $\mu$ is therefore an effective late-time parameter.}

\keywords{$f(R,T)$ gravity, magnetized strange quark matter, MIT bag model, hypersurface-homogeneous cosmology, isotropization, late-time observational constraints}

\maketitle

\section{Introduction}\label{sec:intro}
The expansion of the Universe has been accelerating for the last several billion years. The first direct evidence came
from the luminosity distances of high-redshift type Ia supernovae (SNe Ia)~\citep{Riess1998,Perlmutter1999}. The
Pantheon+ compilation of 1701 light curves of 1550 spectroscopically confirmed SNe Ia~\citep{Scolnic2022} now determines
the matter density of the flat $\Lambda$ cold dark matter ($\Lambda$CDM) model to about 5\% from supernovae
alone~\citep{Brout2022}. The cosmic microwave background (CMB)~\citep{Planck2018VI} and baryon acoustic oscillations (BAO), first
detected in 2005~\citep{Eisenstein2005,Cole2005} and now measured to sub-percent precision by the Dark Energy Spectroscopic
Instrument (DESI)~\citep{DESIDR2}, confirm it independently. Two results have reopened the question of what
drives it. The first is the Hubble tension between the distance-ladder~\citep{Riess2022} and CMB values of the Hubble
constant. The second is the preference for dynamical dark energy, at $2.8$--$4.2\sigma$, found when the BAO of the second DESI data release (DR2) are
combined with CMB and supernova data~\citep{DESIDR2}.

Within general relativity (GR) the acceleration requires a component with negative pressure, of which a cosmological
constant is the simplest and scalar fields the most studied~\citep{CopelandSamiTsujikawa2006}. The alternative is to
modify gravity~\citep{Clifton2012,NojiriOdintsovOikonomou2017,SotiriouFaraoni2010}, for instance through $f(R)$
theories~\citep{Buchdahl1970} or by coupling matter to curvature non-minimally~\citep{Bertolami2007,Bertolami2008}. Its best-known representative is $f(R,T)$ gravity, introduced by Harko
et al.~\citep{Harko2011}, whose Lagrangian depends on the Ricci scalar $R$ and on the trace $T$ of the
energy--momentum tensor; we use its simplest form, $f(R,T)=R+2\mu T$. Such models have been fitted repeatedly to expansion and distance
data~\citep{Nagpal2018EPJC}, among them the linear model~\citep{Myrzakulov2023fRT,Errahmani2025RVE}. Recent fits combine DESI DR2 BAO with Pantheon+ and
expansion-rate data and compare the model with $\Lambda$CDM through information criteria~\citep{Saddiki2026DESI}, or
combine them with chronometer and Union3 supernova data~\citep{VishalTarai2026}. The theory
carries two caveats. The coupling to $T$ makes the matter
energy--momentum tensor non-conserved~\citep{Harko2011,Alvarenga2013}, and the field equations depend on the on-shell
matter Lagrangian, $L_m=p$ or $L_m=-\rho$~\citep{Carvalho2021}. Fisher and Carlson argue that for a separable
$f(R,T)=R+f_2(T)$ the $T$-term can be absorbed into the matter Lagrangian and has no physical significance of its
own~\citep{Fisher2019} (for a reply see Ref.~\citep{HarkoMoraes2020}). Our results do not depend on that reading, but they test a modified
matter--geometry coupling rather than a modified gravitational action. Simple $f(R,T)$ models have also been argued to be in
tension with cosmological data~\citep{VeltenCarames2017}. A recent reassessment finds that power-law and exponential models survive
cosmic-chronometer and stellar-age constraints only near the $\Lambda$CDM limit and should be ruled out as an alternative
to dark energy~\citep{fRTReview}. In the model studied here the acceleration is driven by an effective bag term,
and the coupling is a correction to be constrained by data rather than the cause of the acceleration.

Strange quark matter is a natural candidate for exotic matter in cosmology. Bodmer~\citep{Bodmer1971} and
Witten~\citep{Witten1984} proposed that matter made of up, down and strange quarks may be the true ground state of
strongly interacting matter. Its equation of state is usually taken from the Massachusetts Institute of Technology (MIT) bag model~\citep{Chodos1974,FarhiJaffe1984,Madsen1999},
$p=(\rho-4B_c)/3$ [Eq.~\eqref{eq:MIT}], in which the bag constant $B_c$ acts as a vacuum energy. In the early Universe quark matter existed before the quantum
chromodynamics (QCD) crossover, which lattice calculations place at $T_c\simeq156.5$~MeV~\citep{HotQCD2019}. Very strong
magnetic fields make the quark pressure anisotropic~\citep{IsayevYang2013,Isayev2018}, and weaker
primordial fields, which may have been generated in the early Universe~\citep{Grasso2001,Subramanian2016},
are limited by the CMB to the nanogauss level today~\citep{PlanckPMF2016,BarrowFerreiraSilk1997}.

A homogeneous magnetic field singles out a direction. It is therefore incompatible with a
Friedmann--Lema\^{\i}tre--Robertson--Walker (FLRW) geometry and requires an anisotropic, spatially homogeneous
space-time~\citep{Thorne1967,BarrowFerreiraSilk1997}. The hypersurface-homogeneous space-times~\citep{StewartEllis1968,HajjBoutros1985}
are the simplest such geometries: a single curvature index $k$ selects the locally rotationally symmetric (LRS) Bianchi
type~I model ($k=0$), the Kantowski--Sachs model ($k=+1$)~\citep{KantowskiSachs1966} and the LRS Bianchi type~III model
($k=-1$). Hypersurface-homogeneous models with perfect-fluid sources have also been studied in $f(R,T)$
gravity~\citep{ShaikhKatore2016}, and Bianchi~III models in which the shear is proportional to the
expansion have been fitted to expansion and distance data~\citep{Sarmah2024BIII,Mune2025BIII}. Whatever the geometry, its anisotropy is strongly limited today by the
CMB~\citep{Saadeh2016}. In GR, Bianchi models with a positive cosmological constant approach de~Sitter
space~\citep{Wald1983}; this covers the Bianchi~I and III members, whereas the Kantowski--Sachs member isotropizes only if
the cosmological constant overcomes its positive curvature before recollapse. A viable anisotropic model must therefore reproduce the isotropy of the present Universe, not
merely exhibit anisotropy.

Magnetized strange quark matter (MSQM) cosmologies have been studied mainly in $f(R,T)$ gravity. LRS Bianchi~I models
have been built with linear, bilinear and special forms of the deceleration parameter~\citep{Sahoo2017MSQM,Sahoo2018NewAst},
the first ending in a Big Rip. MSQM has also been studied in Bianchi~I and V~\citep{Aktas2017TJP} and
FLRW~\citep{AktasAygun2017} models, in which the field vanishes in the final solutions. Further models include a time-dependent cosmological
term~\citep{Khalafi2021MSQM} and Bianchi~VI$_0$~\citep{KalkanAktas2021MSQM}, five-dimensional~\citep{KalkanAktas2022MSQM5D}
and inhomogeneous Einstein--Rosen~\citep{Kalkan2022} geometries. Quark matter without a field has been studied in
plane-symmetric~\citep{AgrawalPawar2017}, FLRW~\citep{Nagpal2018} and
Bianchi~I~\citep{Jokweni2023,Singh2023SQM,Thakre2024SQM} models and
compared with expansion-rate data~\citep{Pawar2025ObsQSQM}. In GR, a strange-quark-matter LRS Bianchi~III model closed with the MIT
equation of state, in which the bag constant acts as an effective vacuum energy, has been fitted to chronometer and BAO
data~\citep{Patra2026SQM}. Most of these works share one methodology. The field equations, which are underdetermined for a
magnetized fluid, are closed by kinematic assumptions: a prescribed deceleration parameter or scale factor, often combined
with a fixed ratio $A=B^n$ of the directional scale factors~\citep{Sahoo2017MSQM,Sahoo2018NewAst,AgrawalPawar2017,Jokweni2023,Thakre2024SQM,KalkanAktas2021MSQM,Pawar2025ObsQSQM}.

This methodology leaves four gaps. First, the kinematic assumptions are incompatible with the source. With the
correct Maxwell stress, $A=B^n$ ($n>1$) forces a negative squared field strength in Bianchi~I whenever the volume
expansion accelerates, and a prescribed deceleration parameter
violates the constancy of the bag constant. Moreover, $A=B^n$ holds the mean anisotropy parameter at a constant value, so that
the model cannot isotropize (Sect.~\ref{sec:closure}). Second, the perfect-fluid form of the $f(R,T)$ coupling term does
not apply to the Maxwell field, for which it is minus the field's own energy--momentum tensor~\citep{Harko2011}, so that
the field couples minimally (Sect.~\ref{sec:Lm}). Third, to our knowledge, the dependence of MSQM results on the choice of
matter Lagrangian has not been quantified. Fourth, none of the MSQM models in $f(R,T)$ gravity that we examined has been
fitted to current distance data with their full covariances or compared with $\Lambda$CDM on the same data. It is
therefore not known whether such a model is compatible with the present Universe.

This paper addresses these gaps. We derive the field equations for MIT-bag strange quark matter and a frozen-in magnetic
field in hypersurface-homogeneous space-time, with the Maxwell sector treated exactly and the matter Lagrangian stated,
and show that the $f(R,T)$ coupling acts only on the quark fluid and the dust. We close the system with the MIT equation
of state and Maxwell's equations alone and obtain exact solutions in limiting cases and numerical solutions of the full
system. Such a closure by the source physics has not been used before for MSQM in $f(R,T)$; the closest precedent is the
non-magnetized GR model of Ref.~\citep{Patra2026SQM} mentioned above. For a universe without primordial shear we extend to $f(R,T)$ the general-relativistic
analysis of the anisotropy induced by a homogeneous field~\citep{BarrowFerreiraSilk1997,Barrow1997skew}. We derive how
the coupling changes its radiation-era level and later decay, and, as in Ref.~\citep{BarrowFerreiraSilk1997}, we use the
CMB limit on the present shear to bound the field strength; the same limit bounds the curvature of the
Kantowski--Sachs and Bianchi~III members. We then constrain the model with 32 cosmic-chronometer (CC)
measurements of the expansion rate, 1590 Pantheon+ supernova light curves and the DESI DR2 BAO, and compare it with
$\Lambda$CDM on the same data through information criteria and a Bayes factor. Finally, we propagate the posterior into
the kinematic, dark-energy, magnetic and energy-condition diagnostics, trace the fitted coupling to the dilution of the
dust and interpret the fitted bag term as an effective vacuum energy, as in the GR model of
Ref.~\citep{Patra2026SQM}. We also compare the results with earlier MSQM models and quantify their
dependence on the matter Lagrangian, the supernova sample and the prior.

Section~\ref{sec:fRT} presents $f(R,T)$ gravity with a magnetized quark fluid, including the matter Lagrangian, the
minimal Maxwell coupling and the non-conservation of matter. Section~\ref{sec:geometry} introduces the geometry and the
source and derives the field equations and the restrictions on the parameters. Section~\ref{sec:solutions} contrasts the
physical closure with kinematic ansatzes and gives the exact and numerical solutions, their asymptotics and the isotropy
bounds. Section~\ref{sec:diagnostics} defines the kinematic and geometric diagnostics, the energy conditions and the
sound speed, and Sect.~\ref{sec:obs} confronts the model with observations. Section~\ref{sec:figures} discusses the
best-fit universe, compares it with earlier models, examines the dependence on the matter Lagrangian and states the
limitations. Section~\ref{sec:conclusions} concludes, and Appendix~\ref{app:mcmc} gives the data, likelihood and sampling
details.

\section{$f(R,T)$ gravity with a magnetized quark fluid}\label{sec:fRT}

\subsection{Action and field equations}\label{sec:action}
We use the metric signature $(-,+,+,+)$, units $G=c=1$ (so that $\kappa=8\pi$) and the curvature conventions of
Ref.~\cite{Carroll2004}; an overdot denotes $d/dt$. Following Harko et al.~\cite{Harko2011}, the action of
$f(R,T)$ gravity is
\begin{equation}
S=\frac{1}{16\pi}\int f(R,T)\,\sqrt{-g}\,d^4x+\int L_m\,\sqrt{-g}\,d^4x ,
\label{eq:action}
\end{equation}
where $R$ is the Ricci scalar, $T=g^{ij}T_{ij}$ is the trace of the matter energy--momentum tensor and $L_m$ is the
matter Lagrangian density. For an $L_m$ that depends on the metric but not on its derivatives, the energy--momentum
tensor is
\begin{equation}
T_{ij}=-\frac{2}{\sqrt{-g}}\,\frac{\delta\big(\sqrt{-g}\,L_m\big)}{\delta g^{ij}}
      =g_{ij}L_m-2\,\frac{\partial L_m}{\partial g^{ij}},
\label{eq:Tdef}
\end{equation}
and a perfect fluid has $T_{ij}=(\rho+p)u_iu_j+p\,g_{ij}$ with $u_iu^i=-1$. Varying Eq.~\eqref{eq:action} with respect
to $g^{ij}$ gives the field equations and the tensor $\Theta_{ij}$ derived in Ref.~\cite{Harko2011} (their Eqs.~(11)
and (18)),
\begin{gather}
f_R R_{ij}-\tfrac12 f\,g_{ij}+\big(g_{ij}\Box-\nabla_i\nabla_j\big)f_R=8\pi T_{ij}-f_T\big(T_{ij}+\Theta_{ij}\big),
\label{eq:fieldgeneral}\\
\Theta_{ij}\equiv g^{lm}\,\frac{\delta T_{lm}}{\delta g^{ij}}
=-2T_{ij}+g_{ij}L_m-2g^{lm}\frac{\partial^2L_m}{\partial g^{ij}\,\partial g^{lm}} ,
\label{eq:Theta}
\end{gather}
where $f_R\equiv\partial f/\partial R$ and $f_T\equiv\partial f/\partial T$. We study the simplest and most widely used
model of Ref.~\cite{Harko2011}, $f(R,T)=R+2\mu T$ with constant $\mu$, for which $f_R=1$ and $f_T=2\mu$. The derivative
terms in Eq.~\eqref{eq:fieldgeneral} vanish, and
\begin{equation}
G_{ij}\equiv R_{ij}-\tfrac12Rg_{ij}=8\pi T_{ij}-2\mu T_{ij}-2\mu\,\Theta_{ij}+\mu T\,g_{ij}.
\label{eq:fieldlinear}
\end{equation}
The coupling constant $\mu$ is dimensionless in our units, and $\mu=0$ gives GR.

\subsection{Matter Lagrangian, minimal Maxwell coupling and non-conservation}\label{sec:Lm}
The source is a strange-quark fluid (superscript f) threaded by a magnetic field (superscript em). The two
components have separate Lagrangians, $L_m=L^{(\mathrm f)}+L^{(\mathrm{em})}$, and therefore
$T_{ij}=\Tf_{ij}+\Tem_{ij}$ and $\Theta_{ij}=\Theta^{(\mathrm f)}_{ij}+\Theta^{(\mathrm{em})}_{ij}$.
This separation matters: a magnetized source is \emph{not} a perfect fluid, so the perfect-fluid form of
$\Theta_{ij}$ cannot be applied to the full magnetized tensor.

For the fluid, both $L^{(\mathrm f)}=p$ and $L^{(\mathrm f)}=-\rho$ reproduce the perfect-fluid $T_{ij}$ but lead to
different field equations once matter couples to geometry~\cite{Bertolami2008} (Sect.~\ref{sec:intro}). We adopt
$L^{(\mathrm f)}=p$, with vanishing second metric derivative. This is the choice of the original
formulation~\cite{Harko2011} written in our signature (there $L_m=-p$ in $(+,-,-,-)$), and it allows a direct comparison
with the $f(R,T)$ and MSQM literature. Equation~\eqref{eq:Theta} then gives
\begin{equation}
\Theta^{(\mathrm f)}_{ij}=-2\,\Tf_{ij}+p\,g_{ij}.
\label{eq:Thetaf}
\end{equation}
The alternative $L^{(\mathrm f)}=-\rho$ gives $\Theta^{(\mathrm f)}_{ij}=-2\Tf_{ij}-\rho\,g_{ij}$; its consequences are
recorded below and in Sect.~\ref{sec:litcomp}.

For the electromagnetic field we use Heaviside--Lorentz units and
$L^{(\mathrm{em})}=-\tfrac14F_{ab}F_{cd}\,g^{ac}g^{bd}$, for which Eq.~\eqref{eq:Tdef} gives
\begin{equation}
\Tem_{ij}=F_{ia}F_{j}{}^{a}-\tfrac14 g_{ij}F_{ab}F^{ab},\qquad g^{ij}\Tem_{ij}=0 .
\label{eq:Tem}
\end{equation}
Because $L^{(\mathrm{em})}$ is quadratic in the metric, the last term of Eq.~\eqref{eq:Theta} does not vanish:
$\partial^2L^{(\mathrm{em})}/\partial g^{ij}\partial g^{lm}=-\tfrac12F_{il}F_{jm}$ (symmetrized in $ij$ and in $lm$), and
Eq.~\eqref{eq:Theta} yields
\begin{equation}
\Theta^{(\mathrm{em})}_{ij}=-2\Tem_{ij}+g_{ij}L^{(\mathrm{em})}+F_{il}F_{j}{}^{l}=-\Tem_{ij},
\label{eq:Thetaem}
\end{equation}
the identity stated by Harko et al.~\cite{Harko2011} for the electromagnetic field.

Because the Maxwell tensor is traceless, the trace $T$ comes from the fluid alone,
\begin{equation}
T=g^{ij}T_{ij}=-\rho+3p .
\label{eq:trace}
\end{equation}
Inserting $T_{ij}=\Tf_{ij}+\Tem_{ij}$, Eq.~\eqref{eq:Thetaf} and Eq.~\eqref{eq:Thetaem} into
Eq.~\eqref{eq:fieldlinear} gives
\begin{equation}
G_{ij}=(8\pi+2\mu)\,\Tf_{ij}+8\pi\,\Tem_{ij}+\mu\,(T-2p)\,g_{ij},
\label{eq:FE}
\end{equation}
or, in mixed form with $\Tf{}^i{}_j=\mathrm{diag}(-\rho,p,p,p)$ in comoving coordinates,
\begin{equation}
G^i{}_j=(8\pi+2\mu)\,\Tf{}^{i}{}_{j}+8\pi\,\Tem{}^{i}{}_{j}-\mu\,(\rho-p)\,\delta^i{}_j .
\label{eq:FEmixed}
\end{equation}
Equation~\eqref{eq:FE} is the central result of this section. Three of its features are used throughout the paper.
First, the magnetic field couples \emph{minimally}: the $\mu$-dependence of the Maxwell sector cancels between
$-2\mu\Tem_{ij}$ and $-2\mu\Theta^{(\mathrm{em})}_{ij}$, so the effect of $f(R,T)$ gravity is confined to the quark fluid.
Second, the quark fluid gravitates with effective strength $8\pi+2\mu$, and it acquires an additional isotropic stress
$\mu(T-2p)=-\mu(\rho-p)$. Third, the trace of Eq.~\eqref{eq:FE}, using $g^{ij}G_{ij}=-R$, gives
\begin{equation}
R=-(8\pi+6\mu)\,T+8\mu p .
\label{eq:traceeq}
\end{equation}
For $L^{(\mathrm f)}=-\rho$ the last term of Eq.~\eqref{eq:FE} becomes $\mu(\rho+3p)\,g_{ij}$, and all other terms are
unchanged. Equation~\eqref{eq:FE} agrees with Eq.~(23) of Ref.~\cite{Harko2011} after the change of signature
($g'_{ij}=-g_{ij}$, $T'=-T$); using that expression unchanged with a $(-,+,+,+)$ metric would add a spurious term
$4\mu p\,g_{ij}$.

The contracted Bianchi identity $\nabla^iG_{ij}=0$ applied to Eq.~\eqref{eq:FE} gives
$(8\pi+2\mu)\nabla^i\Tf_{ij}+8\pi\nabla^i\Tem_{ij}+\mu\,\partial_j(T-2p)=0$. For a source-free Maxwell field,
$\nabla^i\Tem_{ij}$ is proportional to $F_{ja}J^a$ and vanishes; in Sect.~\ref{sec:source} this condition fixes the time
dependence of $h^2$ (flux freezing). The fluid therefore obeys
\begin{equation}
(8\pi+2\mu)\,\nabla^i\Tf_{ij}=\mu\,\partial_j(2p-T)=\mu\,\partial_j(\rho-p) ,
\label{eq:nonconservation}
\end{equation}
so matter is conserved only for $\mu=0$ or $\rho-p=$ const. Contracting Eq.~\eqref{eq:nonconservation} with $u^j$, with
$u^j\nabla^i\Tf_{ij}=-[\dot\rho+\theta(\rho+p)]$ and the expansion scalar $\theta=\nabla_iu^i$, gives
\begin{equation}
(8\pi+3\mu)\,\dot\rho-\mu\,\dot p+(8\pi+2\mu)\,\theta\,(\rho+p)=0 .
\label{eq:energy}
\end{equation}
The projection of Eq.~\eqref{eq:nonconservation} orthogonal to $u^i$ equates the four-acceleration of the fluid to the
spatial gradients of $p$ and $\rho-p$. In a spatially homogeneous cosmology these vanish, so comoving observers follow
geodesics and the coupling acts only through the energy exchange of Eq.~\eqref{eq:energy}.

For barotropic matter, $p=w\rho$ with constant $w$, Eq.~\eqref{eq:energy} becomes
$(8\pi+3\mu-\mu w)\,\dot\rho=-(8\pi+2\mu)(1+w)\,\theta\rho$. Using $\theta=\dot V/V$ for the comoving volume $V$,
\begin{equation}
\rho\propto V^{-\gamma(\mu,w)},\qquad
\gamma(\mu,w)=\frac{(8\pi+2\mu)(1+w)}{8\pi+3\mu-\mu w}.
\label{eq:gamma}
\end{equation}
For radiation-like matter ($w=1/3$) this gives $\gamma=(\mu+4\pi)/(\mu+3\pi)$, which corresponds to an effective
equation-of-state parameter $w_{\rm eff}=\gamma-1=\pi/(\mu+3\pi)$; this reduces to $1/3$ when $\mu=0$. Section~\ref{sec:source}
shows that the same exponent governs the quark component $\rho-B_c$ of the MIT bag model. As $\mu\to0$,
Eqs.~\eqref{eq:FE}, \eqref{eq:nonconservation} and \eqref{eq:gamma} reduce to their GR forms.

Two values of the coupling are singular. At $\mu=-4\pi$ the fluid drops out of the tensor part of Eq.~\eqref{eq:FE}
(for $\mu<-4\pi$ it gravitates with negative strength), and at $\mu=-8\pi/(3-w)$ the coefficient of $\dot\rho$ in
Eq.~\eqref{eq:energy} vanishes, so the energy equation degenerates into a constraint. These and the further bounds
derived in Sect.~\ref{sec:fieldeqs} are collected in Table~\ref{tab:restrictions}.

\section{Magnetized strange quark matter in hypersurface-homogeneous space-time}\label{sec:geometry}
\subsection{Geometry and kinematics}\label{sec:kinematics}
We consider the hypersurface-homogeneous space-time
\begin{equation}
ds^2=-dt^2+A^2(t)\,dx^2+B^2(t)\big[dy^2+S_k^2(y)\,dz^2\big],
\label{eq:metric}
\end{equation}
where $t$ is cosmic (proper) time, $A(t)$ and $B(t)$ are the scale factors along the $x$ axis, which will be the
direction of the magnetic field, and across it ($y$, $z$), and $S_k(y)=\sin y,\ y,\ \sinh y$. The curvature index $k$ is fixed by the intrinsic geometry of the
two-surfaces $t,x=$ const. For the two-metric $d\ell^2=dy^2+S^2(y)\,dz^2$ the only non-vanishing Christoffel
symbols are $\Gamma^y{}_{zz}=-SS'$ and $\Gamma^z{}_{yz}=S'/S$. The Ricci tensor is then
$R_{yy}=-S''/S$ and $R_{zz}=-SS''$, so the two-dimensional Ricci scalar is $^{(2)}\!R=-2S''/S$. The Gaussian
curvature ${}^{(2)}\!K={}^{(2)}\!R/2$ is therefore
\begin{equation}
{}^{(2)}\!K=-\frac{S''}{S}=
\begin{cases}+1, & S=\sin y,\\ \ \ 0, & S=y,\\ -1, & S=\sinh y,\end{cases}
\label{eq:kmap}
\end{equation}
i.e., $k=+1\leftrightarrow\sin y$, $k=0\leftrightarrow y$ and $k=-1\leftrightarrow\sinh y$.
With this identification the curvature enters the field equations only through the term $k/B^2$
(Sect.~\ref{sec:fieldeqs}). The case $k=+1$ is the Kantowski--Sachs model~\cite{KantowskiSachs1966}, while $k=0$ and
$k=-1$ are the LRS Bianchi~I and Bianchi~III models. The coordinates $x,y,z$ are
dimensionless, so $A$ and $B$ have the dimension of length; below, lengths are measured in units of the present value $B_0$,
so that $A_0=B_0=1$.

For comoving observers $u^i=\delta^i{}_0$, the directional Hubble rates are $H_x=\dot A/A$ and
$H_y=H_z=\dot B/B$. The spatial volume and mean scale factor are
\begin{equation}
\begin{gathered}
V=AB^2=a^3,\qquad H=\frac{\dot a}{a}=\frac13\big(H_x+2H_y\big),\qquad \theta=3H=\frac{\dot V}{V},\\
q\equiv-\frac{a\ddot a}{\dot a^2}=-1-\frac{\dot H}{H^2},
\end{gathered}
\label{eq:kinematics}
\end{equation}
where $q$ is the deceleration parameter. With $A_0=B_0=1$ also $a_0=1$, and the redshift is $1+z=a_0/a=1/a$.
The shear scalar $\sigma^2=\tfrac12\sigma_{ij}\sigma^{ij}$ and the mean anisotropy parameter follow from
$\sigma^2=\tfrac12\sum_iH_i^2-\theta^2/6$ and $A_m=\tfrac13\sum_i(H_i-H)^2/H^2$. With $H_y=H_z$,
\begin{equation}
\sigma^2=\frac13\big(H_x-H_y\big)^2,\qquad A_m=\frac{2\sigma^2}{3H^2}.
\label{eq:shear}
\end{equation}
The model isotropizes if $\sigma^2/\theta^2\to0$ (equivalently $A_m\to0$).

\subsection{Source: magnetized quark fluid, dust and radiation}\label{sec:source}
The source consists of a strange-quark fluid with energy density $\rho$ and isotropic pressure $p$, together
with a magnetic field. For comoving observers the field has no electric component (shown below). The
magnetic four-vector $h_i$ is spacelike and orthogonal to $u^i$, and is directed along the $x$ axis:
\begin{equation}
u_iu^i=-1,\qquad h_iu^i=0,\qquad h_i=(0,\,A\,h,\,0,\,0),\qquad h^2\equiv h_ih^i\ge0 .
\label{eq:hvec}
\end{equation}
Here $h\ge0$ is the magnitude of the field in geometrized units.
For a purely magnetic field, the Maxwell tensor \eqref{eq:Tem} can be written as
$\Tem_{ij}=h^2\big(u_iu_j+\tfrac12g_{ij}\big)-h_ih_j$. Adding the fluid part
$\Tf_{ij}=(\rho+p)u_iu_j+pg_{ij}$ gives
\begin{equation}
T_{ij}=\big(\rho+p+h^2\big)u_iu_j+\Big(p+\frac{h^2}{2}\Big)g_{ij}-h_ih_j ,
\label{eq:EMT}
\end{equation}
which is the standard energy--momentum tensor of a perfect fluid with a magnetic field~\cite{BarrowMaartensTsagas2007}. Its mixed components are
\begin{equation}
T^i{}_j=\mathrm{diag}\Big(-\big(\rho+\tfrac{h^2}{2}\big),\ p-\tfrac{h^2}{2},\ p+\tfrac{h^2}{2},\ p+\tfrac{h^2}{2}\Big).
\label{eq:EMTmixed}
\end{equation}
Hence
\begin{equation}
\varepsilon=\rho+\frac{h^2}{2},\qquad p_\parallel=p-\frac{h^2}{2}\ \ (\text{along }x),\qquad
p_\perp=p+\frac{h^2}{2}\ \ (\text{in the }yz\text{-plane}),
\label{eq:ppar}
\end{equation}
which expresses the familiar magnetic tension along the field lines and magnetic pressure across them. Because the
magnetic part is traceless,
\begin{equation}
T=T^i{}_i=-\varepsilon+p_\parallel+2p_\perp=-\rho+3p,
\label{eq:traceEMT}
\end{equation}
in agreement with Eq.~\eqref{eq:trace}. The splitting \eqref{eq:EMT} matches the decomposition
$T_{ij}=\Tf_{ij}+\Tem_{ij}$ used in Sect.~\ref{sec:fRT}, so the field equations \eqref{eq:FE} apply with this
source and no further assumption.

A very strong field also makes the \emph{quark-matter} pressure itself anisotropic through Landau quantization. For
cold, dense strange quark matter this happens above $10^{17}\lesssim\Bf_{\rm th}\lesssim10^{18}$~G, and the
longitudinal pressure vanishes near $10^{18}$~G~\cite{IsayevYang2013,Isayev2018}. Assuming that the threshold is not
lower in the hot QCD-epoch plasma, we take $p$ isotropic, so that all pressure anisotropy comes from Eq.~\eqref{eq:ppar};
this is checked against the field strengths of Sect.~\ref{sec:asymptotics}.

In the MIT bag model~\cite{Chodos1974} of strange quark matter (Sect.~\ref{sec:intro}), for massless non-interacting $u$, $d$ and $s$
quarks the pressure and energy density are $p=p_q-B_c$ and $\rho=\rho_q+B_c$, where $p_q=\rho_q/3$ and $B_c$ is the bag
constant. Eliminating $p_q$ gives the equation of state
\begin{equation}
p=\frac13\big(\rho-4B_c\big),\qquad \rho_q\equiv\rho-B_c,\qquad p=\frac{\rho_q}{3}-B_c .
\label{eq:MIT}
\end{equation}
Two consequences are used below: the trace \eqref{eq:traceEMT} is constant,
\begin{equation}
T=-\rho+3p=-4B_c ,
\label{eq:Tconst}
\end{equation}
and $\rho+p=\tfrac43\rho_q$. Inserting these and $\dot p=\dot\rho/3$ into the energy equation \eqref{eq:energy} gives
$\big(8\pi+\tfrac83\mu\big)\dot\rho_q=-\tfrac43(8\pi+2\mu)\,\theta\,\rho_q$, that is,
\begin{equation}
\rho_q\propto V^{-\gamma}=a^{-3\gamma},\qquad \gamma\equiv\gamma\big(\mu,\tfrac13\big)=\frac{\mu+4\pi}{\mu+3\pi},\qquad
w_{q,\mathrm{eff}}=\gamma-1=\frac{\pi}{\mu+3\pi}.
\label{eq:rhoqV}
\end{equation}
The bag constant behaves as a constant vacuum-like energy, while the quark component dilutes like radiation modified
by the coupling $\mu$. For $\mu=0$ one recovers $\rho_q\propto a^{-4}$.

For massless quarks, three-flavor quark matter can be absolutely stable relative to $^{56}$Fe while two-flavor
matter is not, provided $144~\mathrm{MeV}<B_c^{1/4}<163~\mathrm{MeV}$~\cite{FarhiJaffe1984,Madsen1999}. In natural
units, $B_c=(B_c^{1/4})^4/(\hbar c)^3$, so
\begin{equation}
56~\mathrm{MeV\,fm^{-3}}\lesssim B_c\lesssim 92~\mathrm{MeV\,fm^{-3}},\quad\text{i.e.}\quad
B_c\simeq(0.90\text{--}1.47)\times10^{34}~\mathrm{J\,m^{-3}} .
\label{eq:Bcrange}
\end{equation}
This is about $2\times10^{43}$ times the present dark-energy density
$\rho_\Lambda c^2\simeq5.3\times10^{-10}~\mathrm{J\,m^{-3}}$ (for $H_0=67.4$~\kmsMpc{} and $\Omega_\Lambda=0.685$).
In the late-time Universe studied here, where the bag term plays the role of dark energy, $B_c$ (equivalently
$\Lambda_{\rm eff}=(8\pi+4\mu)B_c$ of Sect.~\ref{sec:fieldeqs}, or $\Omega_B$) is therefore an \emph{effective} vacuum
parameter of bag-model form, fixed by observations, not the QCD bag constant. The quark component $\rho_q$ is a
radiation-like remnant whose present abundance $\Omega_{q0}$ is bounded by the data. The physical value
\eqref{eq:Bcrange} applies to the quark epoch: the bag phase ends when $p=\rho_q/3-B_c=0$, i.e., at $\rho=4B_c$, which
for $B_c^{1/4}=144$--$163$~MeV and $\mu=0$ the Friedmann constraint of a quark-plus-bag universe places at
$t=1/(2H)\simeq26$--$33$~\textmu s, consistent with the lattice-QCD crossover temperature
$T_c=156.5\pm1.5$~MeV~\cite{HotQCD2019}. A quantitative treatment of that epoch, including the lepton and photon bath,
is beyond the scope of this paper.

To describe the late Universe we add the two standard components. Baryons and cold dark matter are represented by
a pressureless fluid (dust) with energy density $\rho_d$ and four-velocity $u^i$. With the same prescription as for the
quark fluid, $L^{(d)}=p_d=0$, Eq.~\eqref{eq:Theta} gives $\Theta^{(d)}_{ij}=-2T^{(d)}_{ij}$. The dust therefore enters the
field equations \eqref{eq:FE} through the fluid part, and its trace $-\rho_d$ is added to Eq.~\eqref{eq:Tconst}:
\begin{equation*}
T=-4B_c-\rho_d .
\end{equation*}
The photon background (and, as an approximation, the relativistic neutrinos) is treated as an isotropic superposition
of Maxwell fields, with energy density $\rho_r$ and pressure $\rho_r/3$. By Eq.~\eqref{eq:Thetaem} it couples minimally
and does not contribute to $T$. The non-conservation law \eqref{eq:nonconservation} constrains only the sum of the
fluids. We assume, as usual, that the components do not exchange energy directly, so that each obeys its own version
of Eq.~\eqref{eq:energy}. For dust this gives
$(8\pi+3\mu)\dot\rho_d+(8\pi+2\mu)\theta\rho_d=0$, i.e.,
\begin{equation}
\rho_d\propto V^{-\gamma_d},\qquad \gamma_d=\frac{8\pi+2\mu}{8\pi+3\mu},\qquad w_{d,\rm eff}=\gamma_d-1=-\frac{\mu}{8\pi+3\mu},
\label{eq:rhod}
\end{equation}
while the radiation is conserved, $\rho_r\propto V^{-4/3}$. In linear $f(R,T)$ gravity with ordinary matter alone the
deceleration parameter is constant, and no transition to acceleration occurs~\cite{VeltenCarames2017}. In the present
model the constant MIT trace supplies a constant vacuum term (Sect.~\ref{sec:fieldeqs}), so a transition is
possible.

We take the electromagnetic field as the two-form
\begin{equation}
F=h_0\,S_k(y)\,dy\wedge dz,\qquad F_{yz}=-F_{zy}=h_0\,S_k(y),\qquad h_0=\text{const}.
\label{eq:Fform}
\end{equation}
where the constant $h_0$ fixes the conserved magnetic flux. The homogeneous equations hold identically,
$dF=h_0S_k'\,dy\wedge dy\wedge dz=0$. For the inhomogeneous ones, $\sqrt{-g}=AB^2S_k$ and
$F^{yz}=h_0/(B^4S_k)$ give $\sqrt{-g}\,F^{yz}=h_0A/B^2$, which depends on $t$ only, so
$\nabla_aF^{ab}=(-g)^{-1/2}\partial_a(\sqrt{-g}\,F^{ab})=0$ for all $b$. Both sets of Maxwell equations therefore hold
for \emph{arbitrary} $A(t)$ and $B(t)$, without any current. The electric field measured by comoving observers is $E_a=F_{ab}u^b=0$.
The magnetic vector $h^a=\tfrac12\epsilon^{abcd}u_bF_{cd}$ has only an $x$-component, and
\begin{equation}
h^2=h_ah^a=\tfrac12F_{ab}F^{ab}=\frac{h_0^2}{B^4}.
\label{eq:fluxfreeze}
\end{equation}
This is flux freezing. The magnetic flux through a comoving element of the $yz$ two-surface,
$\int F=h_0\int S_k\,dy\,dz$, is constant, while the field strength decays as $h\propto B^{-2}$, as in the
classical Bianchi~I analyses~\cite{Thorne1967} and in the early-Universe
literature~\cite{Grasso2001}. Equation~\eqref{eq:fluxfreeze} is precisely the condition, found in Sect.~\ref{sec:Lm},
for the Maxwell stress to be separately conserved, $\nabla_i\Tem{}^i{}_j=0$. Hence it is not an extra assumption but a
consequence of Maxwell's equations. (The field \emph{vector} points along $x$, while the two-form $F$ lies in the
$yz$-plane.)

\subsection{Field equations, constraint and restrictions}\label{sec:fieldeqs}
For the metric \eqref{eq:metric} and the source \eqref{eq:EMTmixed} supplemented by dust and radiation, the $(t,t)$, $(x,x)$ and
$(y,y)$ components of Eq.~\eqref{eq:FEmixed} (the $(z,z)$ component repeats the $(y,y)$ one) give, with $\rho=\rho_q+B_c$ and $p=\rho_q/3-B_c$ [Eq.~\eqref{eq:MIT}],
\begin{align}
\frac{2\dot A\dot B}{AB}+\frac{\dot B^2}{B^2}+\frac{k}{B^2}&=\Big(8\pi+\frac{8\mu}{3}\Big)\rho_q+\Lambda_{\rm eff}
+(8\pi+3\mu)\rho_d+8\pi\rho_r+4\pi h^2,\label{eq:F00}\\
\frac{2\ddot B}{B}+\frac{\dot B^2}{B^2}+\frac{k}{B^2}&=-\frac{8\pi}{3}\rho_q+\Lambda_{\rm eff}+\mu\rho_d-\frac{8\pi}{3}\rho_r+4\pi h^2,\label{eq:F11}\\
\frac{\ddot A}{A}+\frac{\ddot B}{B}+\frac{\dot A\dot B}{AB}&=-\frac{8\pi}{3}\rho_q+\Lambda_{\rm eff}+\mu\rho_d-\frac{8\pi}{3}\rho_r-4\pi h^2.\label{eq:F22}
\end{align}
On the right-hand side the fluid part is $(8\pi+2\mu)\,\mathrm{diag}(-\rho,p,p,p)$ with $\rho\to\rho+\rho_d$ for the dust,
the Maxwell part is $8\pi\,\mathrm{diag}(-\tfrac{h^2}{2},-\tfrac{h^2}{2},\tfrac{h^2}{2},\tfrac{h^2}{2})$, the radiation part is
$8\pi\,\mathrm{diag}(-\rho_r,\rho_r/3,\rho_r/3,\rho_r/3)$, and the isotropic term is $-\mu(\rho+\rho_d-p)\delta^i{}_j$. The bag
constant enters all three equations through $\Lambda_{\rm eff}\equiv(8\pi+4\mu)B_c$, which acts exactly as an effective
cosmological constant because the trace $T=-4B_c-\rho_d$ contains $B_c$ only as a constant [Eq.~\eqref{eq:Tconst}].
A similar identification of the coupling term $4\lambda B_c$ with a cosmological constant was made for strange quark
matter without a field in Ref.~\cite{Singh2023SQM}. The
coupling $\mu$ does not enter the quark terms of the dynamical equations \eqref{eq:F11}--\eqref{eq:F22}; it enters only the
constraint \eqref{eq:F00} and, through Eq.~\eqref{eq:rhoqV}, the dilution law of $\rho_q$. The magnetic terms enter with
$+4\pi h^2$ in Eq.~\eqref{eq:F11}, along the field, and with $-4\pi h^2$ in Eq.~\eqref{eq:F22}, across the field; they
carry no factor of $\mu$ (minimal coupling, Sect.~\ref{sec:Lm}). For a diagonal mixed tensor $\mathcal R^i{}_j$ that
depends on $t$ only, the time component of the divergence on the metric \eqref{eq:metric} is
$\nabla_i\mathcal R^i{}_0=\dot{\mathcal R}^0{}_0+\theta\,\mathcal R^0{}_0-H_x\mathcal R^1{}_1-2H_y\mathcal R^2{}_2$. With the
dilution laws \eqref{eq:rhoqV} and \eqref{eq:rhod}, $\rho_r\propto a^{-4}$ and flux freezing \eqref{eq:fluxfreeze}, all terms
in $h_0$ and $B_c$ cancel and the right-hand side satisfies the contracted Bianchi identity identically, so
Eq.~\eqref{eq:F00} is a constraint preserved by Eqs.~\eqref{eq:F11}--\eqref{eq:F22}.

With $H_x=\dot A/A$, $H_y=\dot B/B$ and Eq.~\eqref{eq:shear}, one has $2H_xH_y+H_y^2=3H^2-\sigma^2$. Equation~\eqref{eq:F00}
therefore becomes the generalized Friedmann constraint
\begin{equation}
3H^2=\sigma^2-\frac{k}{B^2}+\Big(8\pi+\frac{8\mu}{3}\Big)\rho_q+\Lambda_{\rm eff}+(8\pi+3\mu)\rho_d+8\pi\rho_r+\frac{4\pi h_0^2}{B^4}.
\label{eq:FriedmannFull}
\end{equation}
Dividing by $3H_0^2$ defines the present-day density parameters (a subscript $0$ denotes the present epoch)
\begin{equation}
\begin{gathered}
\Omega_{q0}=\frac{(8\pi+\frac83\mu)\rho_{q0}}{3H_0^2},\quad
\Omega_{B}=\frac{(8\pi+4\mu)B_c}{3H_0^2},\quad
\Omega_{d0}=\frac{(8\pi+3\mu)\rho_{d0}}{3H_0^2},\quad
\Omega_{r0}=\frac{8\pi\rho_{r0}}{3H_0^2},\\
\Omega_{h0}=\frac{4\pi h_0^2}{3H_0^2B_0^4},\quad
\Omega_{\sigma0}=\frac{\sigma_0^2}{3H_0^2},\quad
\Omega_{k0}=-\frac{k}{3H_0^2B_0^2},
\end{gathered}
\label{eq:Omegas}
\end{equation}
which satisfy $\Omega_{q0}+\Omega_B+\Omega_{d0}+\Omega_{r0}+\Omega_{h0}+\Omega_{\sigma0}+\Omega_{k0}=1$. With $\rho_q>0$ and
$B_c>0$, positivity of the effective quark and vacuum terms requires $\mu>-3\pi$ ($\Omega_q>0$) and $\mu>-2\pi$ ($\Omega_B>0$).
The second condition is new relative to Sect.~\ref{sec:Lm}, and it is the binding condition on the coupling: for
$-8\pi/3<\mu\le-2\pi$ all densities are positive, but the bag term acts as a negative (or vanishing) cosmological constant
and cannot drive acceleration.

Subtracting Eq.~\eqref{eq:F11} from Eq.~\eqref{eq:F22} eliminates the matter, whose terms are identical in both. With
$\Delta\equiv H_x-H_y$ (so $\sigma^2=\Delta^2/3$), and using $\ddot A/A=\dot H_x+H_x^2$ and $\ddot B/B=\dot H_y+H_y^2$, the
left-hand side is $\dot\Delta+(H_x-H_y)(H_x+H_y)+H_y(H_x-H_y)-k/B^2=\dot\Delta+\theta\Delta-k/B^2$. Hence
\begin{equation}
\dot\Delta+\theta\,\Delta=\frac{k}{B^2}-8\pi h^2=\frac{k}{B^2}-\frac{8\pi h_0^2}{B^4},
\qquad
\Delta(t)=\frac{1}{V}\Big[V_{\rm i}\Delta_{\rm i}+\int_{t_{\rm i}}^{t}\Big(\frac{k}{B^2}-\frac{8\pi h_0^2}{B^4}\Big)V\,dt'\Big],
\label{eq:shearEq}
\end{equation}
where the subscript i refers to an arbitrary initial time $t_{\rm i}$. Integrated from the initial singularity, this reads
$V\Delta=\mathcal C+\int_{0}^{t}V\big(k/B^2-8\pi h_0^2/B^4\big)dt'$, where the constant $\mathcal C$ is the free (primordial)
shear. Two consequences follow. First, the magnetic field acts as a \emph{source of anisotropy} with a definite sign: it
drives $H_x<H_y$, so expansion along the field lines is slower than across them, which is the effect of magnetic tension.
Without the source terms, $\Delta\propto V^{-1}$, the familiar free decay of shear. Second, an exactly isotropic solution
($\Delta\equiv0$) requires $k/B^2=8\pi h_0^2/B^4$ at all times; for an expanding model this is possible only if $k=0$ and
$h_0=0$. A frozen-in magnetic field, and likewise non-zero curvature of the $(y,z)$ surfaces, is therefore incompatible with
exact isotropy, in agreement with the classical result for magnetized Bianchi~I models~\cite{Thorne1967}. Whether the
universe \emph{isotropizes asymptotically}, i.e., whether $\sigma^2/\theta^2\to0$, is a dynamical question addressed in
Sect.~\ref{sec:asymptotics}.

Combining Eqs.~\eqref{eq:F11}--\eqref{eq:F22} and eliminating $3H^2$ with Eq.~\eqref{eq:FriedmannFull} (the curvature terms
cancel) gives the acceleration (Raychaudhuri) equation
\begin{equation}
\frac{\ddot a}{a}=-\frac23\sigma^2-\frac49(6\pi+\mu)\rho_q+\frac{\Lambda_{\rm eff}}{3}-\frac{4\pi}{3}\rho_d-\frac{8\pi}{3}\rho_r-\frac{4\pi}{3}h^2 .
\label{eq:RaychFull}
\end{equation}
Shear, quark matter (for $\mu>-6\pi$), dust, radiation and the magnetic field all decelerate the expansion, whereas the bag
term accelerates it when $\mu>-2\pi$. A transition from deceleration to acceleration therefore occurs once the bag term
dominates. For $\mu=0$ the quark, bag and magnetic terms take the form expected for radiation-like quarks
($\rho+3p=2\rho_q$), a cosmological constant $8\pi B_c$ and a magnetic fluid with $\varepsilon+\sum p=h^2$.

Table~\ref{tab:restrictions} collects the conditions on the parameters.

\begin{table}[ht]\centering\small
\caption{Restrictions on the parameters and their origin.}
\label{tab:restrictions}
\begin{tabular}{@{}>{\raggedright\arraybackslash}p{2.8cm}>{\raggedright\arraybackslash}p{3.7cm}>{\raggedright\arraybackslash}p{5.6cm}@{}}\toprule
Condition & Equivalent form & Origin\\\midrule
$8\pi+2\mu>0$ & $\mu>-4\pi$ & positive gravitational strength of the fluids, Eq.~\eqref{eq:FE}\\
$3\pi+\mu>0$ & $\mu>-3\pi$ & non-degenerate quark energy equation; $\Omega_{q0}\ge0$ for $\rho_q\ge0$\\
$8\pi+3\mu>0$ & $\mu>-8\pi/3$ & non-degenerate dust energy equation; $\Omega_{d0}\ge0$ for $\rho_d\ge0$\\
$2\pi+\mu>0$ & $\mu>-2\pi$ & $\Omega_B>0$ for $B_c>0$: bag term accelerates, Eq.~\eqref{eq:RaychFull}\\
$\rho_q,\rho_d,\rho_r\ge0$ & $\Omega_{q0},\Omega_{d0},\Omega_{r0}\ge0$ & physical densities\\
$h^2\ge0$ & $\Omega_{h0}\ge0$ & automatic: $h^2=h_0^2/B^4$\\
real $A,B>0$ & --- & automatic: $A=a\,e^{-2\psi/3}$, $B=a\,e^{\psi/3}$ with real $\psi$\\
$\sum\Omega=1$ & $\Omega_B=1-\Omega_{q0}-\Omega_{d0}-\Omega_{r0}-\Omega_{h0}-\Omega_{\sigma0}-\Omega_{k0}$ & constraint at $z=0$\\
$\mathrm{sign}(\Omega_{k0})=-k$ & $k=+1$: $\Omega_{k0}<0$ & definition, Eq.~\eqref{eq:Omegas}\\
$H^2>0$ for all $a$ & $W>0$ [Eq.~\eqref{eq:Ealg}] & no bounce or recollapse; holds for $|\Omega_{k0}|\ll\Omega_B$\\
field below strong-field threshold & $\Bf\ll10^{17}$ G & isotropic quark pressure (Sect.~\ref{sec:source}); satisfied (Sect.~\ref{sec:asymptotics})\\\bottomrule
\end{tabular}
\end{table}

\section{Closure and solutions}\label{sec:solutions}
\subsection{Physical closure versus kinematic ansatzes}\label{sec:closure}
The general system contains five unknown functions of time, $A$, $B$, $\rho$, $p$ and $h^2$, but only three
independent field equations, \eqref{eq:F00}--\eqref{eq:F22}. Exactly two further conditions are therefore needed.
Any additional condition over-determines the system and, generically, has no solution.

A common practice in the literature is to impose two kinematic conditions: a constant deceleration parameter,
$a\propto(lt+m)^{1/(1+q)}$, and a fixed ratio of directional scale factors, $A=B^n$. These close the system, but they
leave no freedom for the physical equation of state of strange quark matter or for Maxwell's equations. For the
correct magnetized source, $A=B^n$ implies $\Delta=(n-1)H_y$, and the anisotropy equation \eqref{eq:shearEq} gives
\begin{equation}
8\pi h^2=-\frac{3(n-1)}{n+2}\big(\dot H+3H^2\big)+\frac{k}{B^2},\qquad
\dot H+3H^2=\frac{(2-q)\,l^2}{(1+q)^2(lt+m)^2}\quad(q=\text{const}).
\label{eq:hA}
\end{equation}
Since $\dot H+3H^2=\ddot V/(3V)>0$ for any expanding model with $\ddot V>0$, a non-negative $h^2$ requires $n\le1$ for
$k=0$; a model with constant $n>1$ is therefore inconsistent with a real magnetic field. Maxwell's equations further demand
$h^2=h_0^2/B^4\propto(lt+m)^{-12/[(1+q)(n+2)]}$. Comparing powers of $(lt+m)$ in Eq.~\eqref{eq:hA}, the two
expressions can agree for all $t$ only if $k=0$ and $(1+q)(n+2)=6$. With $n<1$ this gives
$q-1=2(1-n)/(n+2)>0$, a strongly decelerating universe. The alternative is $h_0=0$, i.e., no field. In addition, the
MIT condition $B_c=(\rho-3p)/4=\text{const}$ fails: the field equations give $B_c\propto(lt+m)^{-2}$. Replacing the
constant-$q$ law by a prescribed hybrid law $a\propto t^\alpha e^{\beta t}$ does not cure this. With $A=B^n$, the MIT
condition and Maxwell's equations are again violated, because $B_c(t)$ and $h^2B^4$ come out time-dependent.
Moreover, the expansion history is then fixed kinematically by $(\alpha,\beta)$ alone, so data on $H(z)$ cannot
constrain $\mu$ or $B_c$. Finally, with $H_x=nH_y$ Eq.~\eqref{eq:shear} gives a constant anisotropy parameter,
$A_m=2(n-1)^2/(n+2)^2$ (for example $A_m=0.125$ for $n=2$): the proportionality $A=B^n$ used in earlier MSQM
models~\cite{Sahoo2017MSQM,Sahoo2018NewAst} freezes the anisotropy, which then never decays.

We close the system instead with the two conditions that the physics of the source itself provides: the MIT equation of
state \eqref{eq:MIT} and Maxwell's equations, which give flux freezing \eqref{eq:fluxfreeze}. By Eqs.~\eqref{eq:rhoqV} and
\eqref{eq:fluxfreeze}, $\rho_q$ and $h^2$ are then algebraic functions of $V$ and $B$, and the model reduces to the two
evolution equations \eqref{eq:F11}--\eqref{eq:F22} for $A$ and $B$, subject to the constraint \eqref{eq:FriedmannFull}, with
no kinematic ansatz. Once the parameters and initial data satisfying the constraint are given, the model is fully
determined. The expansion history $H(z)$ then depends on the physical parameters
$(H_0,\Omega_{q0},\Omega_{d0},\Omega_{r0},\Omega_B,\Omega_{h0},\Omega_{\sigma0},\Omega_{k0},\mu)$. The coupling $\mu$ enters through
the dilution exponents $3\gamma$ and $3\gamma_d$ and the coefficients of the acceleration equation, so it can in principle be
constrained by observations.

\subsection{Exact solutions and limiting cases}\label{sec:exact}
For $k=0$, $h_0=0$ and $\mathcal C=0$ the model is isotropic. With a single matter component $X$ and the bag term,
Eq.~\eqref{eq:FriedmannFull} gives $H^2=H_0^2(\Omega_{X0}a^{-3\gamma_X}+\Omega_B)$, which integrates to
\begin{equation}
\begin{gathered}
a(t)=\Big(\frac{\Omega_{X0}}{\Omega_B}\Big)^{1/(3\gamma_X)}\sinh^{2/(3\gamma_X)}\!\Big(\tfrac32\gamma_X\sqrt{\Omega_B}\,H_0t\Big),\\
H_0t_0=\frac{2}{3\gamma_X\sqrt{\Omega_B}}\,\mathrm{arsinh}\sqrt{\frac{\Omega_B}{\Omega_{X0}}} ,
\end{gathered}
\label{eq:isoexact}
\end{equation}
with $X=d$ and $\gamma_X=\gamma_d$ for dust plus bag, and $X=q$ and $\gamma_X=\gamma$ for quark matter plus bag. In terms of
the dimensional quantities, $\Omega_B H_0^2=\Lambda_{\rm eff}/3$ and $\Omega_{X0}H_0^2a_0^{3\gamma_X}$ is the coefficient of
$a^{-3\gamma_X}$ in $H^2$. The solution interpolates between a decelerating era, $a\propto t^{2/(3\gamma_X)}$, and a de~Sitter
era, and it serves as the isotropic reference background for the magnetized, anisotropic solutions below. For $\mu\to0$
the field equations reduce to those of GR ($\gamma=4/3$, $\gamma_d=1$, $\Lambda_{\rm eff}=8\pi B_c$); with, in
addition, $\Omega_{q0},\Omega_{h0}\to0$ and $k=0$ the model is flat $\Lambda$CDM with $\Lambda=8\pi B_c$, and
Eq.~\eqref{eq:isoexact} with $X=d$ is its exact scale factor.

For $k=0$, $h_0=0$ and no matter other than the bag term, but with $\mathcal C\neq0$, Eq.~\eqref{eq:FriedmannFull} reduces to
$3H^2=\sigma^2+\Lambda_{\rm eff}$ with $\sigma\propto V^{-1}$. The solution is the exact anisotropic Kasner--de~Sitter
solution, valid for any $\mu$,
\begin{equation}
\begin{gathered}
V=V_\ast\sinh(\omega t),\qquad \omega=\sqrt{3\Lambda_{\rm eff}},\\
A=A_\ast\sinh^{1/3}(\omega t)\,\tanh^{2\varsigma/3}\!\big(\tfrac{\omega t}{2}\big),\qquad
B=B_\ast\sinh^{1/3}(\omega t)\,\tanh^{-\varsigma/3}\!\big(\tfrac{\omega t}{2}\big),
\end{gathered}
\label{eq:kasnerdS}
\end{equation}
with $\varsigma=\pm1$ the sign of $\Delta$. It satisfies all three field equations exactly. As $t\to0$ it approaches the
Kasner solution with $(p_x,p_y)=(1,0)$ for $\varsigma=+1$ or $(-\tfrac13,\tfrac23)$ for $\varsigma=-1$. There the
Kretschmann scalar diverges as $\mathcal K=\tfrac{64}{27}t^{-4}$ for $\varsigma=-1$, a genuine curvature singularity,
whereas for $\varsigma=+1$ it stays finite, $\mathcal K\to4\Lambda_{\rm eff}^2$. As $t\to\infty$ it approaches
de~Sitter space.

Finally, in the general-relativistic limit $\mu=0$ with $k=0$, $h_0=0$ and dust plus bag only ($\Omega_{q0}=\Omega_{r0}=0$),
the volume obeys $\ddot V=3\Lambda_{\rm eff}V+12\pi\rho_dV$, with $\rho_dV$ constant, whose solution is
$V=c_1e^{\omega t}+c_2e^{-\omega t}-4\pi\rho_{d}V/\Lambda_{\rm eff}$, with $\omega=\sqrt{3\Lambda_{\rm eff}}$ as in
Eq.~\eqref{eq:kasnerdS}, where the constraint \eqref{eq:FriedmannFull} fixes
$c_1c_2=\tfrac14\big[(4\pi\rho_dV/\Lambda_{\rm eff})^2-\mathcal C^2/(3\Lambda_{\rm eff})\big]$. In GR, exact Bianchi~III
solutions with a magnetic field, dust and a cosmological constant have been obtained in Ref.~\cite{Zofka2026}.

\subsection{Numerical solution, asymptotics and isotropy bounds}\label{sec:asymptotics}
The densities obey the dilution laws \eqref{eq:rhoqV} and \eqref{eq:rhod} and $\rho_r\propto a^{-4}$, the field obeys flux
freezing \eqref{eq:fluxfreeze}, and the shear obeys the integrated form of Eq.~\eqref{eq:shearEq}; these are first
integrals of the model. We use $\ell=\ln a$ as the independent variable, with the normalization $a_0=A_0=B_0=1$ of
Sect.~\ref{sec:kinematics}. The dependent variables are $\Sigma=\Delta/H$ (so $\sigma^2/\theta^2=\Sigma^2/27$),
$\psi=\ln(B/A)$ and $\tau=H_0t$, with $b\equiv B/B_0=e^{\ell+\psi/3}$ and $E=H/H_0$; then $\Omega_{\sigma0}=\Delta_0^2/(9H_0^2)$.
The constraint \eqref{eq:FriedmannFull} fixes $E$ algebraically:
\begin{equation}
E^2=\frac{W(\ell,\psi)}{1-\Sigma^2/9},\qquad
W=\Omega_{q0}e^{-3\gamma\ell}+\Omega_{d0}e^{-3\gamma_d\ell}+\Omega_{r0}e^{-4\ell}+\Omega_B+\Omega_{h0}b^{-4}+\Omega_{k0}b^{-2}.
\label{eq:Ealg}
\end{equation}
The evolution equations follow from Eqs.~\eqref{eq:RaychFull} and \eqref{eq:shearEq}:
\begin{align}
\frac{d\Sigma}{d\ell}&=\frac{\dot\Delta}{H_0^2E^2}-\Sigma\,\frac{\dot H}{H_0^2E^2},\qquad \frac{d\psi}{d\ell}=-\Sigma,\qquad \frac{d\tau}{d\ell}=\frac1E,\label{eq:dimsys}\\
\frac{\dot H}{H_0^2}&=-E^2\Big(1+\frac{2\Sigma^2}{9}\Big)-\frac{6\pi+\mu}{2(3\pi+\mu)}\Omega_{q0}e^{-3\gamma\ell}-\frac{4\pi\,\Omega_{d0}}{8\pi+3\mu}e^{-3\gamma_d\ell}\nonumber\\
&\quad-\Omega_{r0}e^{-4\ell}+\Omega_B-\Omega_{h0}b^{-4},\nonumber\\
\frac{\dot\Delta}{H_0^2}&=-3E^2\Sigma-3\Omega_{k0}b^{-2}-6\Omega_{h0}b^{-4}.\nonumber
\end{align}
Because the constraint is imposed algebraically, the numerical scheme preserves it exactly. Equations~\eqref{eq:F11}
and \eqref{eq:F22} are not used separately, so they serve as independent checks of the solution.

The free mode $\mathcal C/V$ of the shear grows toward the past faster than any matter component. Unless $\mathcal C$ is
extremely small, the universe is shear dominated before big-bang nucleosynthesis (BBN; see the bound below). As our baseline we
therefore assume \emph{no primordial shear}, $\mathcal C=0$. All anisotropy is then generated by the magnetic field and the
curvature of the $(y,z)$ surfaces, and today's shear $\Sigma_0$ is a \emph{prediction} of the model rather than an input.
Numerically, $\Sigma=0$ is imposed at $a=10^{-20}$ and Eqs.~\eqref{eq:dimsys} are integrated forward, the direction in which
the free mode decays. The initial value of $\psi$ is iterated so that $\psi(0)=0$, and $\Omega_B$ is fixed by $E(0)=1$.
Initial data with $\mathcal C\neq0$ are integrated backward from $z=0$ with a prescribed $\Omega_{\sigma0}$.

\begin{figure}[ht]\centering
\includegraphics[width=\textwidth]{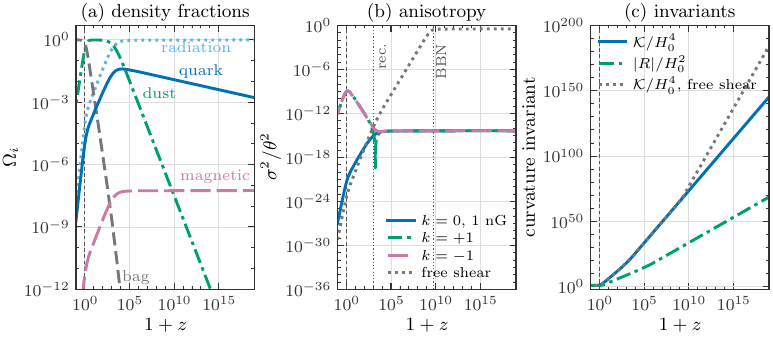}
\caption{History of the best-fit universe from $a=10$ to $a=10^{-19}$, without primordial shear unless stated; the dashed
vertical line marks today. (a) Density fractions $\Omega_i(z)$: successive radiation, dust and bag (effective vacuum)
domination. The quark remnant is radiation-like, and the magnetic fraction stays below $10^{-7}$. (b) Anisotropy
$\sigma^2/\theta^2$ generated by the 1~nG field ($k=0$) and by curvature $\Omega_{k0}=\mp10^{-4}$ ($k=\pm1$), compared with a
free primordial shear $\Omega_{\sigma0}=10^{-23}$. The dip marks a sign change of $\Delta$; the dotted lines mark
recombination (rec.) and BBN ($T=1$~MeV). (c) Kretschmann scalar $\mathcal K/H_0^4$ and Ricci scalar $|R|/H_0^2$, and
$\mathcal K/H_0^4$ for the free primordial shear. Their limits are listed in
Table~\ref{tab:asymptotics}. Best fit of Sect.~\ref{sec:figures}.}
\label{fig:history}
\end{figure}

All numerical results of this section, including Fig.~\ref{fig:history}, are evaluated at the joint observational best fit
defined in Sect.~\ref{sec:figures}. The magnetic density follows from the present
field strength: in Gaussian units a field $\Bf$ corresponds to $h^2=G\,\Bf^2/(4\pi c^4)$ in geometrized units, so that
$\Omega_{h0}=\Bf_0^2/(8\pi\rho_{c0}c^2)$, with $\rho_{c0}$ the present critical density. The system \eqref{eq:dimsys} was
integrated with an explicit eighth-order Runge--Kutta method (DOP853, $r_{\rm tol}=10^{-12}$) from $a=10^{-20}$ to $a=148$,
and the solution was substituted back into \emph{each} of the three field equations \eqref{eq:F00}--\eqref{eq:F22}. The
largest residual, normalized by $H^2$, is $4\times10^{-14}$ at the best fit and for its variants without primordial shear
($-1\le\mu\le2$, $\Omega_{k0}=\pm10^{-4}$, $\Omega_{h0}=0$ and $10^{-10}$), independent of $r_{\rm tol}$ between $10^{-10}$ and
$10^{-13}$. Variants with a free primordial shear, integrated backward from $z=0$, satisfy the field equations to
$10^{-9}$. The exact solutions \eqref{eq:isoexact} and \eqref{eq:kasnerdS} are reproduced to
$3\times10^{-11}$. The symbolic verification of every analytic result of
Sects.~\ref{sec:fRT}--\ref{sec:diagnostics} and the further numerical validation checks will be included in the reproducibility
package, to be released upon publication (see Code availability).

Figure~\ref{fig:history} shows the best-fit universe: (a) the successive radiation, dust and bag domination of the
density fractions, (b) the anisotropy generated by the field and by curvature compared with that of a free primordial
shear, and (c) the curvature invariants. Statements about singularities rest on these invariants: the Ricci scalar of the metric,
$R=2\big[\ddot A/A+2\ddot B/B+2\dot A\dot B/(AB)+\dot B^2/B^2+k/B^2\big]$, which by the trace equation \eqref{eq:traceeq} equals
\[
R=(8\pi+6\mu)(4B_c+\rho_d)+8\mu\Big(\frac{\rho_q}{3}-B_c\Big),
\]
and the Kretschmann scalar,
\[
\mathcal K=4\Big[\Big(\frac{\ddot A}{A}\Big)^2+2\Big(\frac{\ddot B}{B}\Big)^2+2\Big(\frac{\dot A\dot B}{AB}\Big)^2+\Big(\frac{\dot B^2}{B^2}+\frac{k}{B^2}\Big)^2\Big].
\]
Both were derived symbolically from the metric. Table~\ref{tab:asymptotics} gives the early-time, dust-era and late-time
behavior.

\begin{table}[ht]\centering\small
\caption{Asymptotic and era values without primordial shear ($\mathcal C=0$; for $\mu<0$ the quark component dominates at the
earliest times, with the same qualitative behavior except for the sign of $R$). $H_\infty=\sqrt{\Omega_B}H_0=\sqrt{\Lambda_{\rm eff}/3}$. The
diagnostics $j=r$, $s$, $w_{\rm eff}$, $c^2_{s,\rm fl}$ and the energy conditions are defined in Sect.~\ref{sec:diagnostics};
the dust-era values with coupling follow Eq.~\eqref{eq:matterera}. A dash marks quantities without a simple dust-era form.}
\label{tab:asymptotics}
\footnotesize\setlength{\tabcolsep}{4pt}
\begin{tabular}{@{}>{\raggedright\arraybackslash}p{2.3cm}>{\raggedright\arraybackslash}p{3.9cm}>{\raggedright\arraybackslash}p{3.3cm}>{\raggedright\arraybackslash}p{2.7cm}@{}}\toprule
Quantity & early times, $a\to0$ & dust era & late times, $t\to\infty$\\\midrule
$A$, $B$ & $\propto t^{1/2}$ (isotropic radiation era) & $\propto t^{2/(3\gamma_d)}$ & $\propto e^{H_\infty t}$\\
$V$ & $\propto t^{3/2}\to0$ & $\propto t^{2/\gamma_d}$ & $\propto e^{3H_\infty t}$\\
$H$ & $1/(2t)\to\infty$ & $2/(3\gamma_dt)$ & $H_\infty$\\
$q$ & $1$ & $4\pi/(8\pi+3\mu)$ & $-1$\\
$j=r$ & $3$ & $4\pi(16\pi+3\mu)/(8\pi+3\mu)^2$ & $1$\\
$s$ & $4/3$ & $2(\mu+4\pi)/(8\pi+3\mu)$ & $0$\\
$w_{\rm eff}$ & $1/3$ & $-\mu/(8\pi+3\mu)$ & $-1$\\
$\sigma^2/\theta^2$ & plateau $\simeq\tfrac43(\Omega_{h0}/\Omega_{r0})^2$ for $\mu\ge0$, Eq.~\eqref{eq:shearplateau}; $\to0$ for $\mu<0$ & $\propto(1+z)^{2(4-3\gamma_d)}$; $9\times10^{-16}$ at recombination ($\Bf_0=1$ nG, best fit) & $\to0$ ($\propto e^{-6H_\infty t}$ for $k=0$)\\
$\rho_q$, $\rho_d$, $\rho_r$ & $\to\infty$ & $\rho_d$ dominant & $\to0$\\
$p=\rho_q/3-B_c$; $p_\parallel$, $p_\perp$ & $\to+\infty$ & --- & $\to-B_c$\\
$h^2$ & $\propto a^{-4}\to\infty$ & $\propto a^{-4}$ & $\to0$\\
$(h^2/2)/\rho_{\rm fl}$ & $\simeq\Omega_{h0}/\Omega_{r0}$ ($\mu\ge0$) & decreasing & $\to0$\\
$c^2_{s,\rm fl}$ & $1/3$ & $\approx0$ & $\to0$\\
null, weak and dominant energy conditions & satisfied & satisfied & satisfied\\
strong energy condition & satisfied & satisfied & violated\\
$R$ & $|R|\to\infty$ ($R\to-\infty$ for $\mu<0$) & $\propto\rho_d$ & $4\Lambda_{\rm eff}$\\
$\mathcal K$ & $\propto a^{-8}\propto t^{-4}\to\infty$ & $\propto\rho_d^2$ & $24H_\infty^4=8\Lambda_{\rm eff}^2/3$\\\bottomrule
\end{tabular}
\end{table}

The model is not singularity-free: the divergence of $R$ and $\mathcal K$ (Table~\ref{tab:asymptotics}) occurs at a
finite past time, $13.54$~Gyr before the present for the best fit of Sect.~\ref{sec:figures}, and is of isotropic
radiation type when $\mathcal C=0$ and of Kasner type when $\mathcal C\neq0$. In the future the solution approaches
de~Sitter space for all three curvature indices. For $k=0$ and $k=-1$ this is the behavior that the cosmic
no-hair theorem guarantees in GR for Bianchi models with a positive cosmological constant~\cite{Wald1983};
our solutions show that it persists with the $\mu T$ coupling. The Kantowski--Sachs case $k=+1$ lies outside the theorem
and follows from Eq.~\eqref{eq:Ealg}: since $E^2=\Delta^2/(9H_0^2)+W$, the only negative term is $\Omega_{k0}b^{-2}$
($\Omega_{k0}<0$ for $k=+1$), so $H^2\ge H_0^2\big(\Omega_B-|\Omega_{k0}|\,b^{-2}\big)>0$ for all $b\ge1$ whenever
$\Omega_B>|\Omega_{k0}|$. The model then never recollapses, the curvature and shear terms decay relative to the constant
$\Omega_B$, and $H\to H_\infty$. The isotropy bound on $|\Omega_{k0}|$ derived below is about ten orders of magnitude
smaller than $\Omega_B$ (Sect.~\ref{sec:fitresults}), so the condition is met by a wide margin.

Without primordial shear and for $k=0$, the induced anisotropy has a closed form in each era. The shear equation
$\dot\Delta+\theta\Delta=-8\pi h^2$ with $h^2\propto a^{-4}$ gives $\Delta=-8\pi h_0^2a^{-3}\int a^{-1}dt$. In the radiation
era $a\propto t^{1/2}$, so $\int a^{-1}dt=2t/a$, $\Delta=-16\pi h^2t$ and $H=1/2t$. This gives
$\Delta/H=-8\pi h^2/H^2\simeq-6\,\Omega_{h0}/(\Omega_{r0}+\Omega_{q0}a^{4-3\gamma})$, where the quark remnant has been kept in
$H^2$ (with $a\propto t^{1/2}$ this is exact for $\mu=0$), and hence the plateau
\begin{equation}
\frac{\sigma^2}{\theta^2}=\frac{\Delta^2}{27H^2}\simeq\frac43\left(\frac{\Omega_{h0}}{\Omega_{r0}+\Omega_{q0}\,a^{4-3\gamma}}\right)^2\simeq\frac43\left(\frac{\Omega_{h0}}{\Omega_{r0}}\right)^2\quad(\rho_q\ll\rho_r).
\label{eq:shearplateau}
\end{equation}
For $\mu>0$ the remnant fades as $a\to0$ ($4-3\gamma>0$) and the approximation becomes exact; for $\mu=0$ the remnant is
exactly radiation-like and $\Omega_{r0}\to\Omega_{r0}+\Omega_{q0}$; for $\mu<0$ the remnant dominates as $a\to0$ and
$\sigma^2/\theta^2\to0$, so there is no plateau. At the best fit the numerical plateau, $4.27\times10^{-15}$, agrees with the
approximate formula, $4.30\times10^{-15}$, to 1\%. For $\mu=0$ the plateau is the weak-field limit of the
general-relativistic result that the field-induced anisotropy decays only logarithmically during the radiation era, the
ratio of field to radiation energy density falling as $Q/[1+4Q\ln(t/t_0)]$ for a constant
$Q$~\cite{BarrowFerreiraSilk1997,Barrow1997skew}. Here $Q=\Omega_{h0}/\Omega_{r0}$ is of order $10^{-7}$
(Sect.~\ref{sec:res:density}), so the logarithmic decay is negligible. In the matter era $a\propto t^{2/(3\gamma_d)}$ gives
$\Delta/H\propto a^{-(4-3\gamma_d)}$, so $\sigma^2/\theta^2\propto(1+z)^{2(4-3\gamma_d)}$: $(1+z)^2$ for $\mu=0$, in agreement
with the general-relativistic results~\cite{BarrowFerreiraSilk1997,Barrow1997skew}, and $(1+z)^{2.2}$ at the
best fit. The coupling thus enters the shear history only through the dilution exponents $3\gamma$ and $3\gamma_d$; the
late-time decay is given in Table~\ref{tab:asymptotics}.

The model makes definite predictions for today's anisotropy, which can be compared with CMB limits on the present shear.
For the vector mode of the Bianchi~VII$_h$ family, which contains the types VII$_0$, V and I as limits, the limit is
$(\sigma_V/H)_0<4.7\times10^{-11}$ (95\%); the scalar- and tensor-mode limits are weaker, $-6.7\times10^{-11}<(\sigma_S/H)_0<9.6\times10^{-11}$
and $(\sigma_T/H)_0<3.4\times10^{-10}$~\cite{Saadeh2016}. Because these limits were derived for that template family and
not for the LRS shear normalization used here, the comparison is indicative only. A frozen-in field induces a present shear proportional to
$\Bf_0^2$; at the best fit $(\sigma/H)_0=5.8\times10^{-11}\,(\Bf_0/1\,\mathrm{nG})^2$ [Eq.~\eqref{eq:shearfit}]. Applied
literally, the CMB limit therefore bounds a homogeneous field by $\Bf_0\lesssim0.9$~nG
($1.3$~nG with the scalar-mode limit). This is consistent in order of magnitude with the bound
that Barrow, Ferreira and Silk derived for a homogeneous field from the CMB isotropy in the same
way~\cite{BarrowFerreiraSilk1997}. The
often-quoted Planck limit, $B_{1\,\rm Mpc}<4.4$~nG (95\%)~\cite{PlanckPMF2016}, refers to a stochastic field smoothed on
1~Mpc and is quoted here only for orientation. At the best fit, the curvature of the $(y,z)$ surfaces alone induces
$(\sigma/H)_0\simeq0.91\,|\Omega_{k0}|$, so $|\Omega_{k0}|\lesssim5\times10^{-11}$. The Kantowski--Sachs and Bianchi~III
members are thus observationally indistinguishable from $k=0$, which justifies $k=0$ as the main case. A free primordial
shear behaves as $\Omega_\sigma\propto a^{-6}$. Requiring its fraction at $T=1$~MeV, $x/(1+x)$ with $x=\Omega_{\sigma0}a_{\rm BBN}^{-2}/\Omega_{r0}$, to be below $0.1$
gives $\Omega_{\sigma0}<\tfrac19\Omega_{r0}a_{\rm BBN}^2=2.7\times10^{-25}$ at the best fit (numerically, the fraction
is $0.09$ for $\Omega_{\sigma0}=2.5\times10^{-25}$). Here
$a_{\rm BBN}=(4/11)^{1/3}\,T_{\rm CMB}/(1~\mathrm{MeV})$, where the factor $(4/11)^{1/3}$ accounts for the entropy that
electron--positron annihilation later transfers to the photons. This BBN condition is far
stronger than the CMB bound and motivates the choice $\mathcal C=0$. Finally, the physical field strength grows as $B^{-2}$
toward the past. For $\Bf_0=1$~nG it reaches $4.4\times10^{14}$~G at the QCD epoch, more than two orders of magnitude below the
threshold for pressure anisotropy of strange quark matter. The isotropic quark pressure assumed in Sect.~\ref{sec:source}
is therefore justified.

\section{Physical diagnostics}\label{sec:diagnostics}
\subsection{Kinematic and geometric diagnostics}\label{sec:kin}
All quantities below are evaluated analytically from the solution of Sect.~\ref{sec:asymptotics}; no numerical
differentiation is used. The deceleration parameter $q$ of Eq.~\eqref{eq:kinematics} and the jerk parameter $j$ are related by
\begin{equation}
j\equiv\frac{\dddot a}{aH^3}=1+\frac{3\dot H}{H^2}+\frac{\ddot H}{H^3}=q(2q+1)-\frac{dq}{d\ell}.
\label{eq:qj}
\end{equation}
Here $\ddot H$ is the time derivative of Eq.~\eqref{eq:dimsys} along the solution, using $\dot a/a=H$,
$\dot B/B=H_y$ and $d(a^{-n})/dt=-nHa^{-n}$:
\begin{equation*}
\begin{split}
\frac{\ddot H}{H_0^3}={}&-2E\frac{\dot H}{H_0^2}-\frac49\frac{\Delta\dot\Delta}{H_0^3}
+\frac{3\gamma(6\pi+\mu)}{2(3\pi+\mu)}E\,\Omega_{q0}a^{-3\gamma}\\
&+\frac{12\pi\gamma_d}{8\pi+3\mu}E\,\Omega_{d0}a^{-3\gamma_d}
+4E\,\Omega_{r0}a^{-4}+4\frac{H_y}{H_0}\Omega_{h0}b^{-4}.
\end{split}
\end{equation*}
We also use the statefinder pair~\cite{Sahni2003}, the $Om$ diagnostic~\cite{Sahni2008} and the effective
equation-of-state parameter of the total source:
\begin{equation}
r=j,\qquad s=\frac{r-1}{3(q-\tfrac12)},\qquad Om(z)=\frac{E^2(z)-1}{(1+z)^3-1},\qquad w_{\rm eff}=-1-\frac{2\dot H}{3H^2}=\frac{2q-1}{3}.
\label{eq:statefinder}
\end{equation}
In flat $\Lambda$CDM without radiation, $\{r,s\}=\{1,0\}$ and $Om(z)=\Omega_m$ exactly; radiation adds $2\Omega_r(z)$
to $r$. To compare with dark-energy reconstructions that
assume standard matter, we define the apparent dark-energy density $E^2_{\rm DE}=E^2-\Omega_{d0}a^{-3}-\Omega_{r0}a^{-4}$
and its equation of state $w_{\rm DE}=-1-\tfrac13\,d\ln E^2_{\rm DE}/d\ell$.

For an era dominated by a component with constant $w$, Eqs.~\eqref{eq:qj}--\eqref{eq:statefinder} give
$q=\tfrac12(1+3w)$, $j=\tfrac12(1+3w)(2+3w)$ and $s=1+w$. In the dust era of the present model,
$w_{d,\rm eff}=-\mu/(8\pi+3\mu)$ by Eq.~\eqref{eq:rhod}, hence
\begin{equation}
q_{\rm m}=\frac{4\pi}{8\pi+3\mu},\qquad j_{\rm m}=\frac{4\pi(16\pi+3\mu)}{(8\pi+3\mu)^2},\qquad s_{\rm m}=\frac{2(\mu+4\pi)}{8\pi+3\mu}.
\label{eq:matterera}
\end{equation}
The jerk therefore measures the coupling directly: $j_{\rm m}=1$ for $\mu=0$, $1.213$ for $\mu=-1$ and $0.846$ for $\mu=+1$.
For $\mu=0$ the defining ratio of $s_{\rm m}$ is $0/0$ because $q_{\rm m}=\tfrac12$; the value $s_{\rm m}=1$ is its limit $\mu\to0$.
More generally, for any model of the form $E^2=\sum_n\Omega_{n}(1+z)^n$, differentiating twice gives
\begin{equation}
r=1+\sum_n\frac{n(n-3)}{2}\,\Omega_n(z),
\label{eq:rgeneral}
\end{equation}
where $\Omega_n(z)$ is the fractional density of the component with exponent $n$. Pressureless matter ($n=3$) and a
cosmological constant ($n=0$) do not contribute, so $r=1$ in $\Lambda$CDM, and radiation ($n=4$) gives $r>1$. Dust with
$n=3\gamma_d<3$, i.e., $\mu>0$, gives $r<1$; within this model, $r<1$ at late times is therefore a direct signature of $\mu>0$.

\subsection{Energy conditions, sound speed and magnetic weight}\label{sec:EC}
Because the pressure is anisotropic, the energy conditions are formulated with the principal
pressures~\cite{HawkingEllis1973}: the null energy condition (NEC) requires $\varepsilon+p_\parallel\ge0$ and
$\varepsilon+p_\perp\ge0$; the weak energy condition (WEC) adds $\varepsilon\ge0$; the dominant energy condition (DEC)
requires $\varepsilon\ge|p_\parallel|$ and $\varepsilon\ge|p_\perp|$. The strong energy condition (SEC) adds
$\varepsilon+p_\parallel+2p_\perp\ge0$ to the NEC. Here $\varepsilon$ is the total energy density of the source considered, defined below for the physical and the
effective source.
In $f(R,T)$ gravity two sources can be distinguished. The \emph{physical} source consists of the matter fields,
with $\varepsilon=\rho_q+B_c+\rho_d+\rho_r+h^2/2$ and $p_{\parallel,\perp}=\rho_q/3-B_c+\rho_r/3\mp h^2/2$. The \emph{effective}
source is $T^{\rm eff}{}^i{}_j=G^i{}_j/8\pi$, which governs the focusing of geodesics. For the physical source one finds
exactly
\begin{equation}
\begin{gathered}
\varepsilon+p_\parallel=\tfrac43(\rho_q+\rho_r)+\rho_d,\qquad \varepsilon+p_\perp=\tfrac43(\rho_q+\rho_r)+\rho_d+h^2,\\
\varepsilon+p_\parallel+2p_\perp=2(\rho_q+\rho_r)+\rho_d+h^2-2B_c .
\end{gathered}
\label{eq:ECphys}
\end{equation}
Hence the physical NEC and WEC hold identically. The DEC also holds, because $\varepsilon-|p_i|\ge0$ term by term for
$B_c>0$. The SEC fails once the bag term dominates. For the effective source, Eqs.~\eqref{eq:F00}--\eqref{eq:F22} and
\eqref{eq:RaychFull} give
\begin{equation}
\rho^{\rm eff}+p^{\rm eff}_\parallel+2p^{\rm eff}_\perp=-\frac{3}{4\pi}\Big(\frac{\ddot a}{a}+\frac23\sigma^2\Big),
\label{eq:SECeff}
\end{equation}
so the effective SEC is violated precisely when the expansion accelerates (up to the negligible shear term). The
effective NEC, $\rho^{\rm eff}+p^{\rm eff}_\parallel=[(\tfrac{32\pi}{3}+\tfrac{8\mu}{3})\rho_q+(8\pi+2\mu)\rho_d+\tfrac{32\pi}{3}\rho_r]/8\pi\ge0$ for
$\mu>-4\pi$. Since
$2\dot H=-\tfrac13[(E_0-E_1)+2(E_0-E_2)]-2\sigma^2+2k/(3B^2)$, where $E_{0,1,2}$ denote the right-hand sides of
Eqs.~\eqref{eq:F00}--\eqref{eq:F22}, the effective NEC implies $\dot H\le0$ for $k\le0$. For $k=+1$ this holds up to a curvature term that is
negligible by the isotropy bound of Sect.~\ref{sec:asymptotics}. Consequently $w_{\rm eff}\ge-1$: the model has no phantom phase.

Strange quark matter in the MIT bag model has $c_s^2=\partial p/\partial\rho=1/3$, which is both causal and positive. The
adiabatic sound speed of the fluid mixture (quark matter, dust and radiation),
$c_{s,\rm fl}^2=\sum_i\dot p_i/\sum_i\dot\rho_i$, lies between $0$ and $1/3$ at all times. It equals $1/3$ in the radiation era,
falls to $1.4\times10^{-4}$ today at the best fit and tends to zero in the de~Sitter future. The bag constant does not
contribute, since $\dot B_c=0$. The sound speed $c^2_{s,\rm eff}=\dot p_{\rm eff}/\dot\rho_{\rm eff}$ of the effective source,
with $p_{\rm eff}=(p^{\rm eff}_\parallel+2p^{\rm eff}_\perp)/3$, decreases from $1/3$ in the radiation era. For $\mu>0$ it
turns negative once the dust dominates $\dot\rho_{\rm eff}$, at $z\simeq168$ for the best fit, and tends to
$w_{d,\rm eff}=-\mu/(8\pi+3\mu)$, which is $-0.036$ at the best fit; for $\mu=0$ it decreases to zero and stays
non-negative. The negative sign follows from the effective dust pressure, which is $w_{d,\rm eff}$ times the effective dust
density [Eqs.~\eqref{eq:F00}--\eqref{eq:F22}]. It is a property of the effective description, which mixes the matter fields
with the geometric coupling terms, and of the background; it is not a propagation speed. The sound speeds of the physical
fluids stay non-negative: $c_s^2=1/3$ for the quark fluid and $0\le c_{s,\rm fl}^2\le1/3$ for the mixture. These are
background statements, and whether perturbations are stable must be decided by the perturbation equations of $f(R,T)$
gravity (Sect.~\ref{sec:litcomp}).

The ratio of magnetic to fluid energy density is $(h^2/2)/\rho_{\rm fl}$, with $\rho_{\rm fl}=\rho_q+B_c+\rho_d+\rho_r$. Since
$h^2\propto B^{-4}$ and $\rho_r\propto a^{-4}$, it is constant during the radiation era. There it equals
$\Omega_{h0}e^{-4\psi/3}/(\Omega_{r0}+\Omega_{q0})$ for $\mu=0$, where the factor $e^{-4\psi/3}$ differs from unity only by
the anisotropy accumulated since $a=10^{-20}$, and it decreases afterward (values in Sect.~\ref{sec:res:density}). For
fields allowed by the CMB the field is therefore dynamically negligible at all times. Its role is to source the anisotropy
(Sect.~\ref{sec:asymptotics}), not to drive the expansion.

\section{Observational constraints}\label{sec:obs}
In this section the model is confronted with three late-time data sets. The same pipeline is applied to flat
$\Lambda$CDM, so that the two fits can be compared on equal terms. The data sources are listed in
Appendix~\ref{app:mcmc}; the checksums, the chains and the code will be included in the reproducibility package.

\subsection{Distances in an anisotropic background}\label{sec:distances}
Light propagating along different directions of a Bianchi-type metric sees different expansion rates. The angular
dependence of the luminosity distance is controlled by the shear, and to first order the fractional direction
dependence is of the order of the integrated $\sigma/H$ along the line of sight~\citep{Schucker2014}. For the solutions of
Sect.~\ref{sec:solutions} the shear today is negligible (Sect.~\ref{sec:asymptotics}); it grows toward the past only as
$\sigma/H\propto(1+z)^{4-3\gamma_d}$ in the matter era and is bounded by the radiation-era plateau,
$\sigma/H\lesssim2\times10^{-7}$ [Eq.~\eqref{eq:shearplateau}]. All distances are therefore computed from the
mean scale factor $a=V^{1/3}$,
\begin{gather*}
D_M(z)=c\int_0^z\frac{dz'}{H(z')},\qquad D_H(z)=\frac{c}{H(z)},\nonumber\\
D_V(z)=\left[zD_M^2(z)D_H(z)\right]^{1/3},\qquad D_L(z)=(1+z_{\rm hel})\,D_M(z),
\end{gather*}
with $H=\dot a/a=\theta/3$. The error of this approximation was quantified directly. At the best fit, the mean Hubble
rate of the full anisotropic solution (no primordial shear, $\Bf_0=1$~nG) and the isotropic expression below agree to
$3\times10^{-16}$, and the comoving distances agree to $1.4\times10^{-11}$ for $z<2.4$. Both differences are many orders of
magnitude below the data errors ($\gtrsim10^{-3}$).

In the isotropic limit the generalized Friedmann constraint \eqref{eq:FriedmannFull} gives, for $k=0$,
\begin{gather}
E^2(z)\equiv\frac{H^2}{H_0^2}=\Omega_{q0}(1+z)^{3\gamma}+\Omega_{d0}(1+z)^{3\gamma_d}+(\Omega_{r0}+\Omega_{h0})(1+z)^4+\Omega_{B},
\label{eq:E2fit}\\
\Omega_B=1-\Omega_{q0}-\Omega_{d0}-\Omega_{r0}-\Omega_{h0},\nonumber
\end{gather}
with $\gamma$ and $\gamma_d$ from Eqs.~\eqref{eq:rhoqV} and \eqref{eq:rhod}. Each density parameter is the coefficient with
which its component enters Eq.~\eqref{eq:F00} [Eq.~\eqref{eq:Omegas}], multiplied by that component's dilution law, and
flatness ($\Omega_\sigma\simeq0$, $k=0$) fixes the bag term, $\Omega_B=\Lambda_{\rm eff}/(3H_0^2)$. Equation~\eqref{eq:E2fit}
is thus Eq.~\eqref{eq:Ealg} with $\Sigma=0$, $k=0$ and $b=a$. For $\mu=0$, $\Omega_{q0}=\Omega_{h0}=0$, it is exactly flat
$\Lambda$CDM with radiation.

For distances, the coupling $\mu$ has two effects. First, the dust dilutes as $(1+z)^{3\gamma_d}$ rather than
$(1+z)^3$, i.e., it behaves as a fluid with $w_{d,\rm eff}=-\mu/(8\pi+3\mu)$. Second, it re-weights the
$\mu$-dependent quark term. The second effect is unobservable at $z<2.4$ for $\Omega_{q0}\sim10^{-5}$. The data
therefore constrain $\mu$ through the dust dilution exponent.

\subsection{Data and likelihood}\label{sec:data}
\paragraph{Cosmic chronometers.} We use the standard compilation of 32 differential-age measurements of $H(z)$ for
$0.07\le z\le1.965$~\citep{Jimenez2003,Simon2005,Stern2010,Moresco2012,Zhang2014,Moresco2015,Moresco2016,Ratsimbazafy2017,Borghi2022},
as collected in Ref.~\citep{Moresco2022}, with the BC03 values for the measurements of Moresco and collaborators. The
covariance follows the public implementation of the Moresco et al.\ covariance~\citep{Moresco2020}. It contains the
initial-mass-function (IMF) and stellar-population-synthesis (SPS) terms (the stellar-library term is tested in Sect.~\ref{sec:litcomp}),
\begin{equation}
C^{\rm CC}_{ij}=\sigma_i^2\delta_{ij}+\sum_{X\in\{\rm IMF,\ SPS\}}\eta_X(z_i)\,H_{\rm fid}(z_i)\;\eta_X(z_j)\,H_{\rm fid}(z_j),
\label{eq:CCcov}
\end{equation}
where $\eta_X(z)$ are the tabulated fractional systematic errors due to the IMF and to the
SPS model, interpolated linearly in $z$. The systematic terms enlarge
the diagonal errors by up to 41\% and correlate all points. The fractional errors multiply a fiducial
expansion rate $H_{\rm fid}(z)$ (Planck 2018 flat $\Lambda$CDM) rather than the measured values: fully correlated
fractional errors scaled by the data bias a fit low~\citep{DAgostini1994}, here by 1.3~\kmsMpc{}
in $H_0$. The choice of fiducial is immaterial: replacing its $(H_0,\Omega_m)=(67.36,0.3153)$ by $(73.0,0.30)$ or
$(62.0,0.33)$ moves the best-fit $H_0$ of either model by at most $0.13$~\kmsMpc{} and the best-fit $\mu$
by less than $10^{-3}$.

\paragraph{Type Ia supernovae.} We use the Pantheon+ compilation~\citep{Scolnic2022,Brout2022}, denoted SN in data combinations, with its full
statistical and systematic covariance. Following Ref.~\citep{Brout2022}, only the Hubble-flow light curves with
$z_{\rm HD}>0.01$ are kept, where $z_{\rm HD}$ is the CMB-frame redshift corrected for peculiar velocities; this leaves 1590
light curves of 1473 distinct SNe Ia. The Cepheid host distances are not used, so the ten calibrator SNe Ia above this redshift enter as ordinary Hubble-flow objects, as in the standard Pantheon+ likelihood,
and the absolute magnitude $M$ is degenerate with $H_0$. It is marginalized analytically with a flat prior~\citep{Conley2011}:
\begin{equation}
\chi^2_{\rm SN}=\bm r^{\rm T}C^{-1}\bm r-\frac{(\bm 1^{\rm T}C^{-1}\bm r)^2}{\bm 1^{\rm T}C^{-1}\bm 1},\qquad
r_i=m_{b,i}^{\rm corr}-5\log_{10}\frac{D_L(z_i)}{10\,\mathrm{pc}}.
\label{eq:chi2SN}
\end{equation}
This expression follows from completing the square in $M$ in $\chi^2(M)=(\bm r-M\bm1)^{\rm T}C^{-1}(\bm r-M\bm1)$. The
marginal and profile likelihoods therefore coincide up to a constant, and the best-fit magnitude is
$\hat M=\bm1^{\rm T}C^{-1}\bm r/\bm1^{\rm T}C^{-1}\bm1$.

\paragraph{Baryon acoustic oscillations.} We use the DESI DR2 measurements~\citep{DESIDR2} in their Gaussian form,
with the full covariance: $D_V/r_d$ for the Bright Galaxy Survey sample at $z=0.295$, and $D_M/r_d$ and $D_H/r_d$ (12 numbers) for
six tracer bins. These are the luminous-red-galaxy and emission-line-galaxy bins labeled LRG1, LRG2, LRG3+ELG1 and ELG2 in
Ref.~\citep{DESIDR2}, the quasars and the Lyman-$\alpha$ forest. The sound horizon at the drag epoch $r_d$ is treated as a free parameter.
A CMB- or BBN-calibrated $r_d$ would assume the standard pre-recombination expansion history, and with $\mu\neq0$ the dust
and quark components (and hence the sound horizon) evolve differently before recombination. With $r_d$ free, BAO
constrain $\Omega_{d0}$, $\mu$ and the product $H_0r_d$, and $H_0$ itself is fixed by the CC data.

The total number of data points is $N=32+1590+13=1635$ and
$\chi^2=\chi^2_{\rm CC}+\chi^2_{\rm SN}+\chi^2_{\rm BAO}$, with $\chi^2_{\rm X}=\bm r^{\rm T}C_{\rm X}^{-1}\bm r$ for
the CC and BAO residual vectors $\bm r$.

\subsection{Priors and sampling}\label{sec:priors}
The free parameters and priors are listed in Table~\ref{tab:priors}. The radiation density is fixed to
$\Omega_{r0}h^2=4.15\times10^{-5}$ (photons with $T_{\rm CMB}=2.7255$~K~\citep{Fixsen2009} and three massless neutrino
species). The quark remnant is fixed to $\Omega_{q0}=10^{-5}$, because it is radiation-like and invisible at $z<2.4$. The
magnetic density is fixed by $\Bf_0=1$~nG through the conversion of Sect.~\ref{sec:asymptotics}; the data cannot detect it
(Sect.~\ref{sec:fitresults}), and its physical bounds come from the CMB and from isotropy (Sect.~\ref{sec:asymptotics}).
Finally, $k=0$, as for the main case of this paper. The posterior was sampled by Markov chain Monte Carlo (MCMC) with the affine-invariant ensemble code
\textsc{emcee}~\citep{ForemanMackey2013,GoodmanWeare2010}, with differential-evolution and snooker
moves~\citep{terBraak2006,terBraakVrugt2008}, which are efficient for the strongly correlated $(\Omega_{d0},\mu,r_d)$
directions. The sampler settings, the convergence diagnostics and the best-fit search are given in Appendix~\ref{app:mcmc}.
Before the joint fits, the likelihoods were validated against published results: DESI DR2 BAO alone in flat $\Lambda$CDM
gives $\Omega_m=0.2972\pm0.0086$ and $hr_d=101.55\pm0.74$~Mpc, against the published $0.2975\pm0.0086$ and
$101.54\pm0.73$~Mpc~\citep{DESIDR2}, and Pantheon+ alone gives $\Omega_m=0.331\pm0.018$, against
$0.334\pm0.018$~\citep{Brout2022}.

For model selection we use the Akaike information criterion (AIC) and the Bayesian information criterion (BIC),
\begin{equation*}
\mathrm{AIC}=\chi^2_{\min}+2n_{\rm p},\qquad \mathrm{BIC}=\chi^2_{\min}+n_{\rm p}\ln N,
\end{equation*}
with $n_{\rm p}$ the number of fitted parameters ($M$ included)~\citep{Akaike1974,Schwarz1978,Liddle2007}. The differences
$\Delta X=X_{\rm MSQM}-X_{\Lambda\rm CDM}$ are read on the Jeffreys-type scale: $|\Delta|<2$ is no preference, $2$--$6$ is
positive and $>6$ is strong~\citep{KassRaftery1995,Liddle2004}. This scale was derived for the BIC; for the AIC it is a
common convention rather than a calibrated evidence scale.

\begin{table}[ht]\centering\small
\caption{Free parameters and priors (all flat). The dust density parameter $\Omega_{d0}$ is defined in
Eq.~\eqref{eq:E2fit}; for $\mu=0$ it equals $\Omega_{m0}$.}
\label{tab:priors}
\begin{tabular}{@{}ll>{\raggedright\arraybackslash}p{6.0cm}@{}}\toprule
Parameter & Prior & Justification\\\midrule
$H_0$ [\kmsMpc] & $\mathcal U(50,90)$ & covers Planck and SH0ES values by many $\sigma$; set by CC only\\
$\Omega_{d0}$ & $\mathcal U(0.05,0.70)$ & physically positive dust, wide compared to the posterior\\
$\mu$ (MSQM only) & $\mathcal U(-6,6)$ & inside the physical bound $\mu>-2\pi$ (Table~\ref{tab:restrictions}); $w_{d,\rm eff}$ spans $[-0.14,\,0.84]$
over the prior, far wider than the posterior\\
$r_d$ [Mpc] & $\mathcal U(100,200)$ & free sound horizon; no early-Universe prior (see text)\\
$M$ & flat, analytic & marginalized in Eq.~\eqref{eq:chi2SN}\\\midrule
fixed & \multicolumn{2}{l}{$\Omega_{r0}h^2=4.15\times10^{-5}$, $\Omega_{q0}=10^{-5}$, $\Bf_0=1$ nG,}\\
 & \multicolumn{2}{l}{$k=0$, no primordial shear}\\\bottomrule
\end{tabular}
\end{table}

\subsection{Constraints and model selection}\label{sec:fitresults}
\begin{figure}[t]\centering
\includegraphics[width=\textwidth]{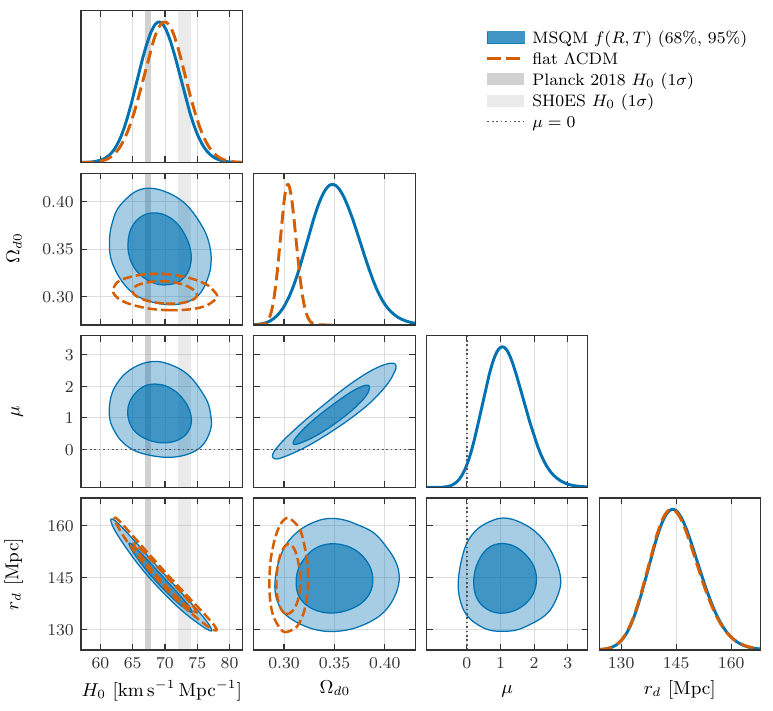}
\caption{Joint posterior from CC+SN+BAO for the MSQM model (filled, 68\% and 95\%) and flat $\Lambda$CDM (dashed). The gray
bands in the $H_0$ column are the Planck 2018 (darker) and SH0ES $1\sigma$ intervals of $H_0$; the dotted lines mark $\mu=0$.
Priors and fixed parameters: Table~\ref{tab:priors}.}
\label{fig:triangle}
\end{figure}

Table~\ref{tab:fit} lists the best fits and the marginalized 68\% and 95\% intervals for the joint data set CC+SN+BAO, and
Fig.~\ref{fig:triangle} shows the joint posterior. The marginals are close to Gaussian, and the $H_0$--$r_d$ ellipse is the
BAO constraint on $H_0r_d$. The Hubble constant is set by the CC data and is discussed in Sect.~\ref{sec:res:expansion}.
The coupling is found to be
\begin{equation*}
\mu=1.12^{+0.65}_{-0.58}\ (68\%),\qquad 0.03<\mu<2.47\ (95\%),\qquad P(\mu>0)=0.98 .
\end{equation*}
This value lies well inside the physical domain $\mu>-2\pi$ of Table~\ref{tab:restrictions}. It is an $O(1)$ number in units
where $\kappa=8\pi$, but small in the sense that $|\mu|/8\pi\simeq0.04$ is the fractional strength of the $T$-dependent
correction to the matter coupling. It corresponds to a dust component with a small negative effective equation of state,
$w_{d,\rm eff}=-\mu/(8\pi+3\mu)=-0.039\pm0.019$. Such dust dilutes slightly more slowly than $a^{-3}$, because energy
flows from the $f(R,T)$ coupling into the matter sector [Eq.~\eqref{eq:rhod}].

The parameters $\Omega_{d0}$ and $\mu$ form a narrow, positively correlated ridge in Fig.~\ref{fig:triangle} (correlation coefficient
$0.93$), and $\mu=0$ lies at the lower edge of the 95\% region: the data determine the combination of matter
content and coupling, not each separately. The reason is that a larger $\mu$ makes the dust dilute more slowly, so a larger
present dust fraction is needed to reproduce the same distances at $z\sim1$. The MSQM $\Omega_{d0}$ is therefore larger than
the $\Lambda$CDM $\Omega_{m0}$, but the physical present dust density,
$\rho_{d0}/\rho_{c0}=8\pi\Omega_{d0}/(8\pi+3\mu)=0.308\pm0.008$, agrees with it and is consistent with the Planck matter density
$0.315\pm0.007$~\citep{Planck2018VI} (a consistency check rather than a test, since Planck assumes standard dilution). The
sound horizon is $r_d=145\pm7$~Mpc with $hr_d=100.0\pm0.8$~Mpc; the $\Lambda$CDM value is $hr_d=101.0\pm0.7$~Mpc.
The values of $H_0$ and $r_d$ are almost identical in the two models.

\begin{table}[ht]\centering\small
\caption{Joint constraints (CC+SN+BAO). Best fit, median with 68\% interval, and 95\% interval. The row $M$ gives the best-fit
Pantheon+ absolute magnitude (marginalized in the sampling). The unit of $H_0$ is \kmsMpc; $\Omega_{d0}=\Omega_{m0}$ in
$\Lambda$CDM.}
\label{tab:fit}
\setlength{\tabcolsep}{2.5pt}\footnotesize
\begin{tabular}{@{}lccc|ccc@{}}\toprule
 & \multicolumn{3}{c|}{MSQM $f(R,T)$} & \multicolumn{3}{c}{flat $\Lambda$CDM}\\
Parameter & best fit & 68\% & 95\% & best fit & 68\% & 95\%\\\midrule
$H_0$ & $69.36$ & $69.12^{+3.27}_{-3.21}$ & $[62.88,\,75.56]$ & $70.09$ & $69.94^{+3.24}_{-3.25}$ & $[63.53,\,76.31]$\\
$\Omega_{d0}$ & $0.3453$ & $0.3495^{+0.0260}_{-0.0243}$ & $[0.3028,\,0.4018]$ & $0.3040$ & $0.3042^{+0.0080}_{-0.0076}$ & $[0.2892,\,0.3200]$\\
$\mu$ & $1.01$ & $1.12^{+0.65}_{-0.58}$ & $[0.03,\,2.47]$ & $0$ (fixed) & --- & ---\\
$r_d$ [Mpc] & $144.3$ & $144.6^{+6.9}_{-6.4}$ & $[132.6,\,158.7]$ & $144.1$ & $144.4^{+7.0}_{-6.3}$ & $[132.5,\,158.8]$\\
$hr_d$ [Mpc] & --- & $99.96^{+0.83}_{-0.82}$ & $[98.36,\,101.61]$ & --- & $101.00^{+0.66}_{-0.67}$ & $[99.69,\,102.30]$\\
$M$ (best fit) & $-19.3724$ & --- & --- & $-19.3574$ & --- & ---\\
$\chi^2_{\rm CC,SN,BAO}$ & \multicolumn{3}{c|}{14.45, 1402.76, 10.01} & \multicolumn{3}{c}{14.42, 1405.28, 10.90}\\
\bottomrule
\end{tabular}
\end{table}

\begin{figure}[t]\centering
\includegraphics[width=\textwidth]{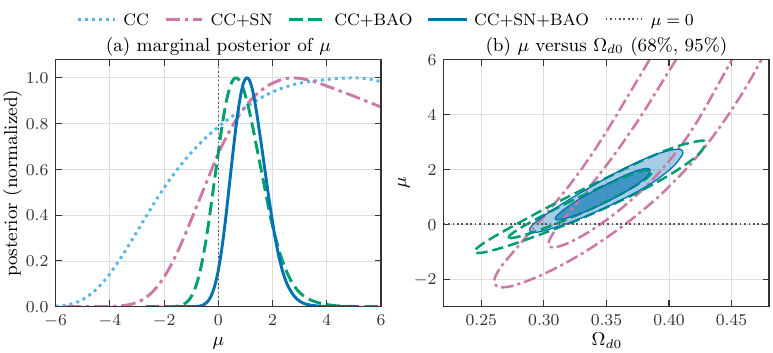}
\caption{Robustness of the coupling. (a) Marginal posterior of $\mu$ for the four data combinations; CC alone is limited by
the prior $|\mu|<6$. (b) $\Omega_{d0}$--$\mu$ contours (68\% and 95\%) for CC+SN, CC+BAO and CC+SN+BAO; CC alone is prior-limited and
appears only in panel (a), and the CC+SN contours are truncated by the prior edge $\mu=6$. SN and BAO give different slopes of the $\Omega_{d0}$--$\mu$ degeneracy, and their intersection drives the joint
result (Table~\ref{tab:fit}). Dotted lines: $\mu=0$.}
\label{fig:robust}
\end{figure}

The four data combinations in Fig.~\ref{fig:robust} show where the signal comes from. CC alone constrains $\mu$ only
weakly; its posterior runs into the prior edge $\mu=6$. CC+SN and CC+BAO each prefer $\mu>0$ only mildly, with $P(\mu>0)=0.86$ and
$0.85$, but their $\Omega_{d0}$--$\mu$ degeneracies have different slopes, and the joint posterior selects their
intersection. The intersection is non-trivial because, within flat $\Lambda$CDM, Pantheon+ and DESI DR2 prefer somewhat
different matter densities (Sect.~\ref{sec:priors}).
The coupling relaxes this difference: SN data, which measure the shape of $D_L(z)$, prefer the slower dust dilution, while
BAO, which measure $D_M/r_d$ and $D_H/r_d$, then fix $\Omega_{d0}$.

\begin{figure}[t]\centering
\includegraphics[width=\textwidth]{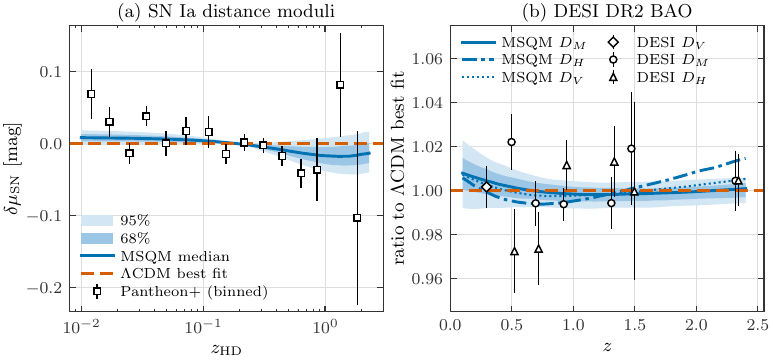}
\caption{Distance data relative to the flat $\Lambda$CDM best fit (dashed; Table~\ref{tab:fit}). (a) Binned Pantheon+
residuals of the 1590 SN light curves of Sect.~\ref{sec:data} (inverse-covariance weights within each bin) and the MSQM
prediction $\delta\mu_{\rm SN}=5\log_{10}(D_M^{\rm MSQM}/D_M^{\Lambda\rm CDM})+\hat M_{\rm MSQM}-\hat M_{\Lambda\rm CDM}$.
(b) DESI DR2 $D_V/r_d$, $D_M/r_d$ and $D_H/r_d$ divided by the $\Lambda$CDM prediction, and the corresponding MSQM ratios, each
model with its own $r_d$; $D_M$ and $D_H$ points at the same redshift are offset slightly for clarity. The MSQM curves are
posterior medians, with 68\% and 95\% bands for $\delta\mu_{\rm SN}$ and $D_M/r_d$. Distances use the mean scale factor
$a=V^{1/3}$ (Sect.~\ref{sec:distances}).}
\label{fig:SNBAO}
\end{figure}

Figure~\ref{fig:SNBAO} compares the model with the supernova and BAO distances and shows where the $\Delta\chi^2$ of
Table~\ref{tab:comparison} comes from. The MSQM curve in Fig.~\ref{fig:SNBAO}(a) lies $0.008$~mag above $\Lambda$CDM at $z=0.01$, crosses zero at
$z\simeq0.18$ and reaches $-0.018$~mag at $z\simeq1.3$, following the trend of the binned residuals at
$0.3<z<1$. In Fig.~\ref{fig:SNBAO}(b), $D_M/r_d$ stays within 0.2\% of $\Lambda$CDM for $z>0.5$, while $D_H/r_d$
dips by 0.6\% near $z\simeq0.76$ and rises by 1.4\% at $z=2.33$. All differences are below the
size of the individual error bars, so the preference for $\mu>0$ is a collective effect of many points. Pantheon+ prefers
distances slightly shorter at $z\sim1$ than a $\Lambda$CDM model calibrated at low $z$; in standard cosmology this is read as
a higher $\Omega_m$. In the MSQM model the same shape is produced by dust that dilutes more slowly, while BAO keep the
absolute $D_M/r_d$ fixed. The same difference between the preferred matter densities contributes to the DESI preference for
dynamical dark energy discussed in Sect.~\ref{sec:res:DE}.

Table~\ref{tab:comparison} compares the goodness of fit of the two models. The joint data set
gives $\chi^2_{\min}=1427.21$ for the MSQM model and $1430.60$ for $\Lambda$CDM, i.e., $\Delta\chi^2=-3.39$ for one extra
parameter. Both fits are statistically acceptable, with $\chi^2_{\rm red}=0.876$ and $0.877$. The value
$\chi^2_{\rm red}<1$ comes from the Pantheon+ covariance rather than from either model: SN alone give
$\chi^2/N\simeq0.88$ for both. The information criteria give
\begin{equation*}
\Delta\mathrm{AIC}=-1.39,\qquad \Delta\mathrm{BIC}=+4.01 .
\end{equation*}
AIC therefore shows no significant preference ($|\Delta|<2$). BIC, which penalizes the extra parameter by $\ln N=7.4$,
gives positive evidence \emph{for} $\Lambda$CDM. For the sub-combinations, $\Delta\chi^2$ is between $-0.6$ and
$0$; AIC shows no preference and BIC favors $\Lambda$CDM. Because the two models are nested at $\mu=0$ and the
prior on $\mu$ is flat, the Bayes factor follows from the Savage--Dickey density ratio,
$B_{01}=p(\mu=0\,|\,d)/\pi(\mu=0)$~\citep{Trotta2008}. It is $B_{01}=1.3$ for the prior $\mathcal U(-6,6)$, which is
inconclusive ($|\ln B_{01}|<1$); its prior dependence is discussed in Sect.~\ref{sec:litcomp}. We conclude that current late-time data are fully compatible with the MSQM model and
with a non-zero $f(R,T)$ coupling of order unity, but do not \emph{require} it. The mild preference for $\mu>0$ comes
entirely from combining SN and BAO, and it should be revisited with future data releases.

\begin{table}[ht]\centering\small
\caption{Goodness of fit and model comparison. The number of fitted parameters, $n_{\rm p}$, includes $M$ when SN data are used,
and $\Delta X=X_{\rm MSQM}-X_{\Lambda\rm CDM}$.}
\label{tab:comparison}
\footnotesize\setlength{\tabcolsep}{3.5pt}
\begin{tabular}{@{}lcrrrrrrrr@{}}\toprule
Data & $N$ & \multicolumn{2}{c}{$\chi^2_{\min}$} & \multicolumn{2}{c}{$\chi^2_{\rm red}$} & $\Delta\chi^2$ & $\Delta$AIC & $\Delta$BIC & $P(\mu>0)$\\
 & & MSQM & $\Lambda$CDM & MSQM & $\Lambda$CDM & & & & \\\midrule
CC+SN+BAO & 1635 & 1427.21 & 1430.60 & 0.8756 & 0.8771 & $-3.39$ & $-1.39$ & $+4.01$ & 0.98\\
CC+SN & 1622 & 1416.98 & 1417.45 & 0.8758 & 0.8755 & $-0.47$ & $+1.53$ & $+6.92$ & 0.86\\
CC+BAO & 45 & 24.14 & 24.73 & 0.5889 & 0.5889 & $-0.59$ & $+1.41$ & $+3.22$ & 0.85\\
CC & 32 & 14.41 & 14.42 & 0.4970 & 0.4805 & $0.00$ & $+2.00$ & $+3.46$ & 0.73\\
\bottomrule
\end{tabular}
\end{table}

With $\Omega_B=0.650\pm0.025$, the bag term that the data require is
\begin{equation}
B_c=\frac{3H_0^2c^2\,\Omega_B}{(8\pi+4\mu)\,G}=(4.4\pm0.7)\times10^{-10}\ \mathrm{J\,m^{-3}},\qquad
B_c^{1/4}=2.15\pm0.09\ \mathrm{meV},
\label{eq:Bcfit}
\end{equation}
which is $0.85\pm0.07$ times the dark-energy density $\rho_\Lambda c^2=\Omega_B\rho_{c0}c^2$. The QCD bag constant of
Eq.~\eqref{eq:Bcrange} is $\sim10^{11}$ times larger in $B_c^{1/4}$ and
$\sim10^{43}$ times larger in $B_c$. The data therefore push $B_c$ to a value that is unphysical for QCD: the late-time
bag term cannot be the confinement energy of strange quark matter. This was anticipated by the effective-bag interpretation of
Sect.~\ref{sec:source}, in which the QCD bag constant governs only the quark--hadron epoch. The same reading of
the bag constant as an effective vacuum energy was obtained in GR for a strange-quark-matter LRS Bianchi~III
model~\citep{Patra2026SQM}.

The magnetic density enters the distances as a radiation-like term,
$\Omega_{h0}(1+z)^4$. At $z<2.4$ it is degenerate with $\Omega_{d0}$ and $\mu$. Profiling the joint likelihood over the
other parameters gives only the weak bound
\begin{equation}
\Omega_{h0}<3\times10^{-2}\ (95\%),\qquad \Bf_0<8\times10^{4}\ \mathrm{nG}.
\label{eq:Ohbound}
\end{equation}
This bound is about $4\times10^{8}$ times weaker in $\Omega_{h0}$ (about $2\times10^{4}$ times in field strength) than the CMB
bounds of a few nG and much weaker than the isotropy bound, both given in Sect.~\ref{sec:asymptotics}. Along the profile, $\mu$ grows with $\Omega_{h0}$: the
additional radiation-like term is compensated by slower dust dilution. The bound is reached where $\mu$ meets the prior
edge $\mu=6$, so it holds only for that prior. The profile has a shallow minimum at $\Omega_{h0}\simeq0.02$,
where $\mu\simeq4.2$ and $\Delta\chi^2=-1.16$ relative to $\Omega_{h0}=0$. The bound lies
$\Delta\chi^2=3.84$ above this minimum and is interpolated linearly between profile nodes spaced by $0.01$, so
it is quoted to one significant figure. We therefore keep $\Bf_0=1$~nG. The posterior then gives
$\Omega_{h0}=(4.9\pm0.4)\times10^{-12}$ (the spread comes only from $H_0$), and a derived present shear
\begin{equation}
\Omega_{\sigma0}=(1.12^{+0.24}_{-0.19})\times10^{-21},\qquad (\sigma/H)_0=(5.8\pm0.6)\times10^{-11}.
\label{eq:shearfit}
\end{equation}
This confirms a posteriori that distances may be computed from the mean scale factor (Sect.~\ref{sec:distances}).

\section{Results and discussion: the best-fit universe}\label{sec:figures}
This section discusses the evolution of the observationally calibrated model. Unless a caption states otherwise, every
curve uses the joint best fit of Table~\ref{tab:fit} with the fixed inputs of Table~\ref{tab:priors}, for which
$\Omega_{r0}=8.63\times10^{-5}$, $\Omega_{h0}=4.90\times10^{-12}$ and, from flatness, $\Omega_B=0.6546$. The 68\% and 95\% bands are
obtained by propagating 300 points drawn from the joint posterior through the anisotropic solver of
Sect.~\ref{sec:asymptotics} and the diagnostics of Sect.~\ref{sec:diagnostics} and taking the percentiles at each redshift.
Sensitivity figures vary one parameter at a time and keep all others at the best fit. The plotted numbers will be tabulated in the
reproducibility package. Table~\ref{tab:derived} gives the posterior of the derived quantities.

\begin{table}[ht]\centering\small
\caption{Posterior of derived quantities (median and 68\% interval; CC+SN+BAO), from posterior samples passed through the
full anisotropic solver with $\Bf_0=1$~nG and no primordial shear (Appendix~\ref{app:mcmc}). The energy conditions are checked
over the complete history $10^{-20}\le a\le e^{3}$. The effective SEC fails exactly at $z_t$ [Eq.~\eqref{eq:SECeff}]. The
derived parameters of the fit ($\Omega_B$, $B_c$, $w_{d,\rm eff}$, the physical dust density, $\Omega_{h0}$ and the present
shear) are given in Sect.~\ref{sec:fitresults}.}
\label{tab:derived}
\begin{tabular}{@{}lcc@{}}\toprule
Quantity & MSQM $f(R,T)$ & flat $\Lambda$CDM\\\midrule
$q_0$ & $-0.496^{+0.027}_{-0.026}$ & $-0.544^{+0.013}_{-0.011}$\\
$z_t$ & $0.648\pm0.022$ & $0.660^{+0.020}_{-0.022}$\\
$j_0$ & $0.940^{+0.029}_{-0.031}$ & $1$\\
$t_0$ [Gyr] & $13.55\pm0.60$ & $13.42^{+0.63}_{-0.61}$\\
$w_{\rm eff}(0)$ & $-0.664^{+0.018}_{-0.017}$ & $-0.696^{+0.008}_{-0.007}$\\
$z^{\rm phys}_{\rm SEC}$ & $0.556^{+0.053}_{-0.056}$ & $0.660^{+0.020}_{-0.022}$\\
$w_{\rm DE}(0)$ (std.\ dust) & $-1.021^{+0.011}_{-0.013}$ & $-1$\\
$w_{\rm DE}(1)$ (std.\ dust) & $-1.75^{+0.45}_{-0.97}$ & $-1$\\
$j$ (dust era) & $0.830^{+0.082}_{-0.077}$ & $1$\\
NEC, WEC, DEC (phys.\ and eff.) & all 1500 samples & all 1500 samples\\
\bottomrule
\end{tabular}
\end{table}

\subsection{Expansion history, deceleration and the present Universe}\label{sec:res:expansion}
\begin{figure}[t]\centering
\includegraphics[width=\textwidth]{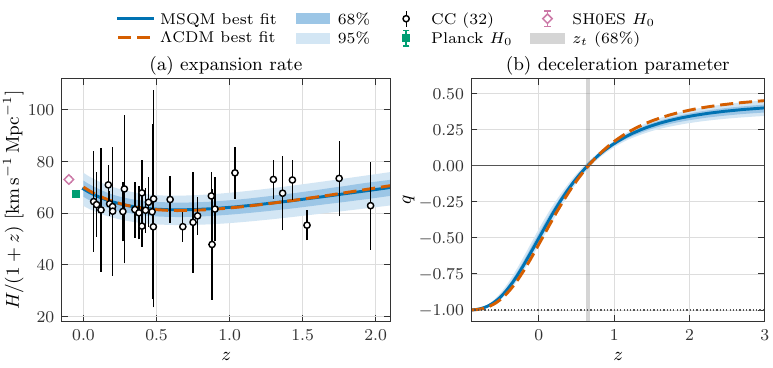}
\caption{Expansion history. (a) $H(z)/(1+z)$ with the 32 CC points (statistical and systematic errors; Sect.~\ref{sec:data})
and the Planck 2018 and SH0ES values of $H_0$, plotted at $z<0$ for visibility. (b) Deceleration parameter for
$-0.9\le z\le3$, with the 68\% interval of the transition redshift $z_t$ (gray). Solid: MSQM best fit (Sect.~\ref{sec:figures})
with 68\% and 95\% posterior bands; dashed: flat $\Lambda$CDM best fit (Table~\ref{tab:fit}).}
\label{fig:expansion}
\end{figure}

Figure~\ref{fig:expansion} shows the expansion history of the best-fit universe. Panel (a) plots the expansion speed
$H/(1+z)=\dot a$ against the 32 CC points. It falls from $H_0$ to a minimum near $z\simeq0.6$--$0.7$ and rises
again at higher redshift; because $d\dot a/dt=\ddot a$, the minimum is the point $q=0$ of Eq.~\eqref{eq:kinematics}, the
deceleration--acceleration transition. Panel (b) plots $q(z)$ itself for $-0.9\le z\le3$. Without any imposed kinematic law, $q$ rises with $z$ toward the
matter-era plateau $q_{\rm m}=0.446$ of Eq.~\eqref{eq:matterera}, slightly below the $\Lambda$CDM value $1/2$; the
plateau is reached only at $z\gg3$, outside the panel, and $q\to1$ in the radiation era (Table~\ref{tab:asymptotics}). It crosses zero at $t\simeq0.56\,t_0$ and tends to $-1$ without crossing it, so there is no
super-acceleration (Sect.~\ref{sec:EC}). The MSQM and $\Lambda$CDM best
fits are visually indistinguishable in panel (a), where their ratio deviates from unity by at most $1.0\%$ over
the plotted range ($2.2\%$ at $z=3$). The 68\% band has a half-width of $4.5\%$ at $z=0$ and is set by the $H_0$ uncertainty of the
CC data. In panel (b) the MSQM curve is shallower than the $\Lambda$CDM one: less negative today and slightly less
decelerating at $z>1$. This is the imprint of $\mu>0$. Dust that dilutes as $(1+z)^{3\gamma_d}$ [Eq.~\eqref{eq:rhod}], with
$3\gamma_d=2.89$ at the best fit, makes the matter era less decelerating, and the bag term then takes over
earlier (Table~\ref{tab:cpl}).

The Hubble constant of Table~\ref{tab:fit} differs from the Planck
value $67.4\pm0.5$~\kmsMpc~\citep{Planck2018VI} by $0.5\sigma$ ($0.8\sigma$ for
$\Lambda$CDM) and lies $1.2\sigma$ ($0.9\sigma$) below the SH0ES value
$73.04\pm1.04$~\kmsMpc~\citep{Riess2022}. With their present errors the CC data cannot arbitrate the Hubble
tension, and the model neither creates nor resolves it: a coupling $\mu\sim1$ changes $H(z)$ only at the percent level,
degenerately with $\Omega_{d0}$. The derived kinematics of Table~\ref{tab:derived} agree with the present Universe. The transition redshift lies within $1\sigma$ of $0.64^{+0.12}_{-0.09}$~\citep{Haridasu2018} and of
$0.59^{+0.12}_{-0.11}$ (from $H(z)$) and $0.683^{+0.11}_{-0.082}$ (from SNe Ia)~\citep{Jesus2020}; it is $1.3\sigma$
below $0.72\pm0.05$ (from $H(z)$, with $H_0=68$~\kmsMpc)~\citep{Farooq2017} and $2.4\sigma$ above the early estimate
$0.4\pm0.1$~\citep{Moresco2016}. The deceleration parameter $q_0$, less negative than in $\Lambda$CDM, lies
$0.4\sigma$ from the SH0ES cosmographic value $-0.51\pm0.024$ obtained in their fit including high-$z$
SNe Ia~\citep{Riess2022} and close to the Planck $\Lambda$CDM value of Table~\ref{tab:compare}. The age differs by
$0.4\sigma$ from the Planck value $13.797\pm0.023$~Gyr~\citep{Planck2018VI}. The late-time expansion is thus that of a matter-plus-vacuum universe, as observed, in which the quark
remnant and the magnetic field are invisible. In this respect the curves resemble $\Lambda$CDM, not the earlier MSQM models,
whose $q(t)$ is prescribed (Sect.~\ref{sec:litcomp}).

\subsection{Dark-energy phenomenology: statefinder, $Om$ and the equivalent $w_0w_a$ model}\label{sec:res:DE}
\begin{figure}[t]\centering
\includegraphics[width=\textwidth]{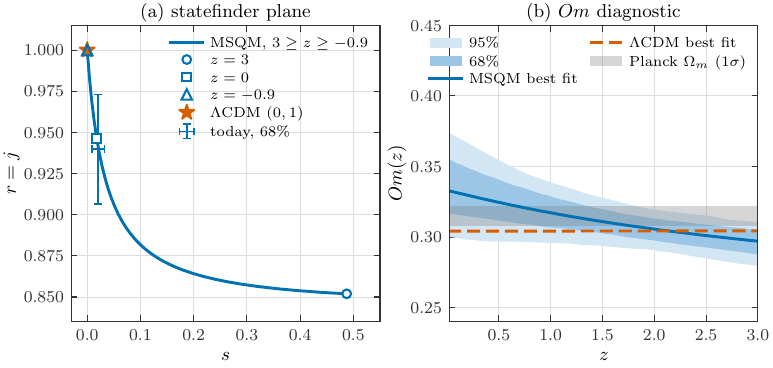}
\caption{Geometric diagnostics. (a) Trajectory in the statefinder plane $\{s,r\}$ from $z=3$ ($\circ$) through today
($\square$, with the 68\% posterior interval) to $z=-0.9$ (open $\triangle$, which nearly coincides with the $\Lambda$CDM fixed point $(0,1)$, $\star$).
(b) $Om(z)$ [Eq.~\eqref{eq:statefinder}] with 68\% and 95\% posterior bands, the flat $\Lambda$CDM best fit (dashed; there
$Om=\Omega_{m0}$) and the Planck 2018 $1\sigma$ band of $\Omega_m$. Best fit of Sect.~\ref{sec:figures}.}
\label{fig:sf}
\end{figure}

Figure~\ref{fig:sf} shows the two geometric diagnostics of Eq.~\eqref{eq:statefinder}. In panel (a) the statefinder
trajectory starts at $(s,r)\simeq(0.49,0.85)$ at $z=3$, passes $s_0=0.018$ today (with $r_0=j_0$ of
Table~\ref{tab:cpl}) and moves monotonically toward the $\Lambda$CDM fixed point $\{r,s\}=\{1,0\}$~\citep{Sahni2003}, which
it reaches in the de~Sitter future. It stays in the quadrant $r<1$, $s>0$, the quintessence-like region of the statefinder
classification~\citep{Sahni2003}. The reason is Eq.~\eqref{eq:rgeneral}: dust with exponent $n=3\gamma_d<3$ contributes
$r-1=-0.16\,\Omega_d(z)$ at the best fit, so $r<1$ is the signature of $\mu>0$, and the posterior $j_0$
(Table~\ref{tab:derived}) lies $2.0\sigma$ below the $\Lambda$CDM value $j=1$. Panel (b) shows that $Om(z)$ decreases
from $0.333$ at $z\to0$ through $0.325$ at $z=0.5$ to $0.306$ at $z=2$, crossing the Planck $1\sigma$ band of
$\Omega_m$~\citep{Planck2018VI}, whereas in $\Lambda$CDM it is constant and equal to $\Omega_{m0}$. By Eq.~\eqref{eq:E2fit} the
decrease follows from the dust exponent $3\gamma_d<3$. Since $Om(z)$ needs no assumed $\Omega_m$ and, for dark energy with
constant $w$, decreases with $z$ when $w>-1$ and increases when $w<-1$~\citep{Sahni2008}, it classifies the MSQM expansion
as quintessence-like, in agreement with the statefinder.

\begin{figure}[t]\centering
\includegraphics[width=\textwidth]{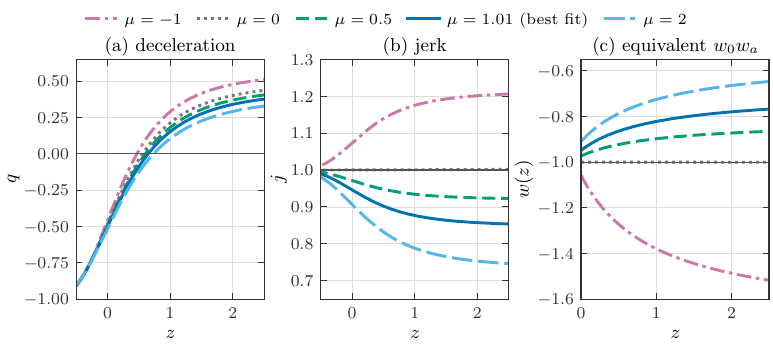}
\caption{Effect of varying only $\mu$, with all other parameters at the best fit of Sect.~\ref{sec:figures}. (a) $q(z)$;
(b) jerk $j(z)$; (c) equation of state $w(z)$ of the equivalent flat $w_0w_a$CDM model of
Table~\ref{tab:cpl}. For $\mu=0$ the curves reduce to $j=1$ and $w=-1$; note that $\mu=0$ with the
best-fit $\Omega_{d0}$ is not the $\Lambda$CDM best fit.}
\label{fig:musens}
\end{figure}

Figure~\ref{fig:musens} shows how $q$, $j$ and the equation of state $w(z)=w_0+w_az/(1+z)$ of the equivalent flat $w_0w_a$CDM model respond
when only $\mu$ is varied at fixed $\Omega_{d0}$; Table~\ref{tab:cpl} lists the corresponding parameters. A larger
$\mu$ strengthens the effect described in Sect.~\ref{sec:res:expansion}: the matter era decelerates less, so $q$ and the
matter-era jerk of Eq.~\eqref{eq:matterera} fall. It also lasts longer in cosmic time, so $t_0$ and $z_t$ increase; this is the
origin of the $\Omega_{d0}$--$\mu$ degeneracy of Sect.~\ref{sec:fitresults}. Panel (c) shows how a standard $w_0w_a$CDM
analysis with free $\Omega_m$ would interpret the MSQM universe. The equivalent model reproduces the MSQM $H(z)$ and $D_M(z)$ to the
residuals of Table~\ref{tab:cpl}, far below the current errors, and it moves away from $(w_0,w_a)=(-1,0)$, the point
reached at $\mu=0$, in a direction set by the sign of $\mu$. For $\mu>0$ it reads the slowly diluting dust as a smaller
$\Omega_m$ plus quintessence-like dark energy with $w_0>-1$ and $w_a>0$; for $\mu<0$ the dark energy is phantom-like, with
$w_0<-1$ and $w_a<0$.

\begin{table}[ht]\centering\small
\caption{One-parameter dependence on $\mu$ (all other parameters at the best fit of Sect.~\ref{sec:figures}) and the equivalent flat $w_0w_a$CDM model
that reproduces the MSQM $H(z)$ and $D_M(z)$ for $0.01<z<2.4$.}
\label{tab:cpl}
\begin{tabular}{@{}rcccccccc@{}}\toprule
$\mu$ & $q_0$ & $z_t$ & $j_0$ & $t_0$ [Gyr] & $\Omega_m^{w_0w_a}$ & $w_0$ & $w_a$ & max.\ resid.\\\midrule
$-1.00$ & $-0.458$ & $0.468$ & $1.074$ & $12.49$ & $0.417$ & $-1.058$ & $-0.643$ & $0.34\%$\\
$0.00$ & $-0.482$ & $0.559$ & $1.000$ & $13.06$ & $0.345$ & $-1.000$ & $0.000$ & $0.00\%$\\
$+0.50$ & $-0.492$ & $0.603$ & $0.972$ & $13.31$ & $0.315$ & $-0.973$ & $+0.151$ & $0.07\%$\\
$+1.01$ & $-0.500$ & $0.648$ & $0.946$ & $13.54$ & $0.288$ & $-0.948$ & $+0.253$ & $0.11\%$\\
$+2.00$ & $-0.515$ & $0.734$ & $0.907$ & $13.95$ & $0.243$ & $-0.909$ & $+0.368$ & $0.15\%$\\
\bottomrule
\end{tabular}
\end{table}

The quintessence-like reading requires $\Omega_m$ to be free. A reconstruction that assumes standard dust with
$\Omega_{m0}=\Omega_{d0}$ reads the same expansion history as phantom-like, with $w_{\rm DE}(z=1)<-1$ (Table~\ref{tab:derived}),
although the total equation of state never crosses $-1$ (Sect.~\ref{sec:EC}); the apparent phantom behavior is an artifact
of fixing $\Omega_m$. DESI DR2 BAO combined with CMB and SN data prefer dynamical dark energy in the quadrant $w_0>-1$,
$w_a<0$, with a phantom crossing in the past~\citep{DESIDR2}. The best-fit MSQM universe (best-fit row of
Table~\ref{tab:cpl}) shares the sign of $w_0+1$ but has $w_a>0$ and never crosses $w=-1$, so a constant coupling cannot
mimic the DESI-preferred evolution. In the MSQM model, the late-time freedom that $w_0w_a$
analyses assign to dark energy is carried by the matter sector instead (Sect.~\ref{sec:fitresults}).

\subsection{Matter sector and the effective bag constant}\label{sec:res:density}
\begin{figure}[t]\centering
\includegraphics[width=\textwidth]{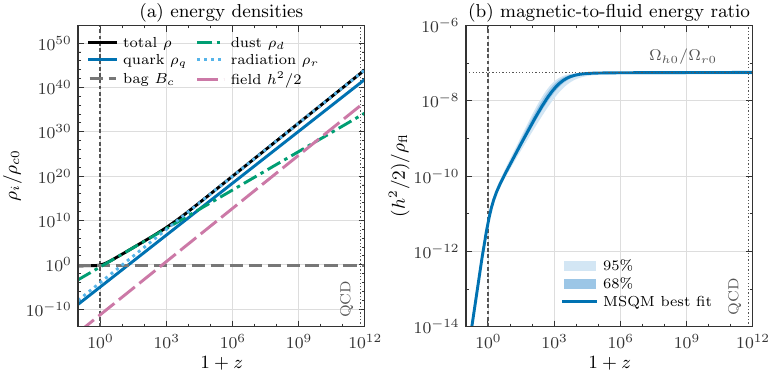}
\caption{Matter sector and magnetic field from $a=10$ ($1+z=0.1$) to $1+z=10^{12}$, just beyond the QCD crossover
(dotted line, $1+z\simeq6.6\times10^{11}$); the dashed line marks today. (a) Energy densities in units of the present critical density
$\rho_{c0}=3H_0^2/8\pi$: total $\rho$, the MIT quark part $\rho_q$, the bag constant $B_c$, dust $\rho_d$, radiation $\rho_r$
and the field energy $h^2/2$ for $\Bf_0=1$~nG. (b) Magnetic-to-fluid energy ratio $(h^2/2)/\rho_{\rm fl}$
($\rho_{\rm fl}$ = quark + bag + dust + radiation) with 68\% and 95\% posterior bands; the dotted line is the
radiation-era value $\Omega_{h0}/\Omega_{r0}$. Best fit of Sect.~\ref{sec:figures}.}
\label{fig:matter}
\end{figure}

Figure~\ref{fig:matter} shows the matter sector and the magnetic field from $a=10$ back to the QCD epoch. In panel (a) every
energy density is positive and decreases monotonically with time, except the constant $B_c$, which overtakes the dust at
$z\simeq0.23$, compared with matter--$\Lambda$ equality at $z\simeq0.29$ in the Planck
$\Lambda$CDM model~\citep{Planck2018VI}. The slopes are those of Eqs.~\eqref{eq:rhoqV}, \eqref{eq:rhod} and
\eqref{eq:fluxfreeze}: $\rho_q\propto(1+z)^{3\gamma}$ with $3\gamma=3.90$ at the best fit,
$\rho_d\propto(1+z)^{3\gamma_d}$, and $\rho_r,\,h^2\propto(1+z)^4$. Since $3\gamma$ is slightly below 4, the
quark-to-radiation ratio decreases slowly into the past, from $\rho_q/\rho_r=0.10$ today to $0.007$
at the QCD epoch. That epoch is governed by the physical bag constant of Eq.~\eqref{eq:Bcrange}, not by the late-time
effective $B_c$ plotted here, which plays the role of dark energy [Eq.~\eqref{eq:Bcfit}]. As in the observed Universe, the
physical dust density today is that of Planck (Sect.~\ref{sec:fitresults}). The pressure-to-density ratio moves through
three stages, $+1/3$ in the radiation era, $\simeq0$ in the matter era and $-1$ under bag domination, with
$p_\parallel/\rho=-0.647$ today; the negative pressure that drives the acceleration is the MIT bag pressure $-B_c$
of Eq.~\eqref{eq:MIT}. Panel (b) shows the magnetic-to-fluid energy ratio. It is constant, $\Omega_{h0}/\Omega_{r0}=5.7\times10^{-8}$,
during the radiation era, because field and radiation dilute alike (Sect.~\ref{sec:EC}); it falls as $(1+z)^{4-3\gamma_d}$ in
the matter era ($\simeq(1+z)^{1.1}$ at the best fit) and steeply once $B_c$ dominates. The field splits the
two pressures by $p_\perp-p_\parallel=h^2$ [Eq.~\eqref{eq:ppar}], twice this ratio. The field therefore cannot affect
the expansion, and the late-time data give only the weak bound \eqref{eq:Ohbound}. Figure~\ref{fig:matter} is thus the thermal history of the standard model with two MSQM features,
the negative bag pressure and a weak magnetic anisotropy of the stress. The anisotropic pressure, which earlier MSQM papers
treat as the defining feature of the model, is real but dynamically negligible at all epochs for a field allowed by the
CMB.

\subsection{Magnetic field and isotropization}\label{sec:res:magnetic}
Flux freezing alone [Eq.~\eqref{eq:fluxfreeze}] gives $\Bf^2=\Bf_0^2(1+z)^4e^{-4\psi/3}$, where the anisotropic
correction $e^{-4\psi/3}-1$ is at most $1.7\times10^{-5}$ (as $a\to10^{-20}$) and $9\times10^{-6}$ at the QCD epoch, where the field is
far below the strong-field threshold of strange quark matter (Sect.~\ref{sec:asymptotics}).

\begin{figure}[t]\centering
\includegraphics[width=\textwidth]{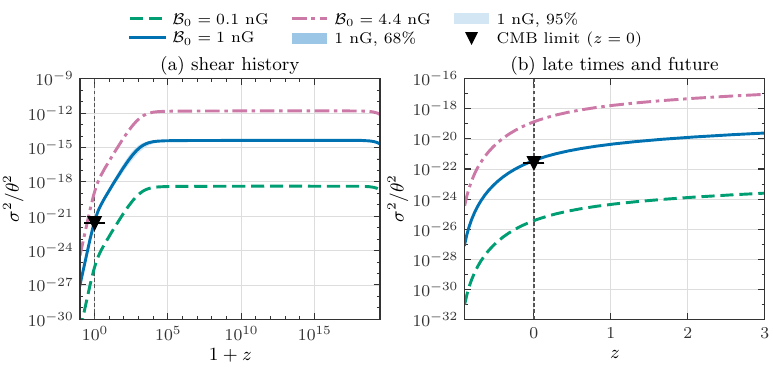}
\caption{Shear history for no primordial shear ($V\Delta=0$ at $a=10^{-20}$), varying only the present field
$\Bf_0\in\{0.1,1,4.4\}$~nG. (a) $\sigma^2/\theta^2$ from $a=3\times10^{-20}$ to $a=10$, with the 68\% and 95\% posterior bands
for $\Bf_0=1$~nG. (b) Late times and future. The triangle on the dashed line (today) is the CMB limit
$(\sigma/H)_0<4.7\times10^{-11}$~\citep{Saadeh2016}, which applies only at $z=0$, expressed as $\sigma^2/\theta^2=(\sigma/H)^2/9$.
Other parameters: best fit of Sect.~\ref{sec:figures}.}
\label{fig:shear}
\end{figure}

Figure~\ref{fig:shear} shows the shear history $\sigma^2/\theta^2$ without primordial shear for three present field
strengths. After a transient of a few e-folds from the initial condition, the curves settle on the radiation-era plateau of
Eq.~\eqref{eq:shearplateau}, the weak-field limit of the logarithmic decay known from GR (Sect.~\ref{sec:asymptotics}).
They then fall as $(1+z)^{2(4-3\gamma_d)}$ once matter dominates and decay faster still under bag domination, so that the model isotropizes, as derived in Sect.~\ref{sec:asymptotics}. Because the field is the only source
of anisotropy [Eq.~\eqref{eq:shearEq}], all curves scale as $\Bf_0^4$: the ratio of the 4.4 and 1~nG curves is
$374.81\simeq4.4^4$. Today $\sigma^2/\theta^2=(\sigma/H)_0^2/9$ with $(\sigma/H)_0$ of Eq.~\eqref{eq:shearfit}, which is of
the order of the CMB vector-mode limit~\citep{Saadeh2016} marked at $z=0$; the indicative bound on $\Bf_0$ that follows is
given in Sect.~\ref{sec:asymptotics}. Together with the Planck bound on a stochastic field~\citep{PlanckPMF2016}, it places
the field at the nanogauss level or below, so the model universe is isotropic today to $10^{-10}$, as the observed Universe
is. The history follows the no-hair behavior of Sect.~\ref{sec:asymptotics} and differs qualitatively from the earlier MSQM models with $A=B^n$, whose anisotropy parameter is
frozen at a constant value (Sect.~\ref{sec:closure}).

\subsection{Energy conditions}\label{sec:res:EC}
\begin{figure}[t]\centering
\includegraphics[width=\textwidth]{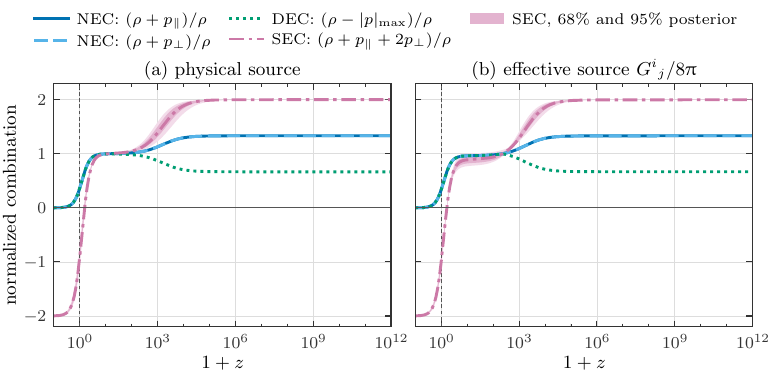}
\caption{Energy conditions normalized by the energy density: $(\rho+p_\parallel)/\rho$, $(\rho+p_\perp)/\rho$ (NEC),
$(\rho-\max|p|)/\rho$ (DEC, with WEC implied) and $(\rho+p_\parallel+2p_\perp)/\rho$ (SEC, with its 68\% and 95\% posterior
bands), from $a=10$ to $1+z=10^{12}$; the dashed line marks today. (a) Physical source (quark + bag + dust + radiation +
field). (b) Effective source $T^{\rm eff}_{ij}=G_{ij}/8\pi$, which includes the $\mu T$ coupling terms. Onset of SEC violation: $z_t$ in panel (b) and
$z^{\rm phys}_{\rm SEC}$ in panel (a) (Table~\ref{tab:derived}). Best fit of Sect.~\ref{sec:figures}.}
\label{fig:EC}
\end{figure}

Figure~\ref{fig:EC} shows the energy conditions, normalized by the energy density, for the physical source (a) and for the
effective source $G_{ij}/8\pi$ (b). The NEC, WEC and DEC stay positive and approach zero only asymptotically in the
vacuum-dominated future (minimum $7\times10^{-4}$ at $a=10$), as Eq.~\eqref{eq:ECphys} requires; the two NEC curves
differ only by its magnetic term $h^2/\rho$. The SEC combination falls from $2$ in the radiation era to $1$ in the
matter era and then turns negative and tends to $-2$, because it contains $-2B_c$. The effective SEC lies below the physical
one in the matter era ($0.9$ instead of 1), which is the effect of $w_{d,\rm eff}<0$, and by
Eq.~\eqref{eq:SECeff} it changes sign exactly at $z_t$. The physical SEC fails later, at the $z^{\rm phys}_{\rm SEC}$ of
Table~\ref{tab:derived}, whereas in $\Lambda$CDM the two redshifts coincide. The interval between them, in which the physical
source still satisfies the SEC while the geometry already accelerates, measures the energy transferred to the dust by the
$\mu T$ coupling. Over the posterior the NEC, WEC and DEC hold over the entire history for both sources
(Table~\ref{tab:derived}). The model therefore has no phantom matter (NEC), a positive energy density (WEC) and causal
energy flow (DEC). Its SEC violation is not a defect but the condition for acceleration that the observations require,
exactly as in $\Lambda$CDM, where the SEC fails below the transition redshift of Table~\ref{tab:compare}.

\subsection{Comparison with earlier models, matter-Lagrangian dependence and limitations}\label{sec:litcomp}
Table~\ref{tab:compare} compares the present results with flat $\Lambda$CDM (our pipeline and Planck 2018), with the
SH0ES cosmography and with six earlier quark-matter cosmologies in $f(R,T)$ gravity. Four differences stand out. The first
is the method. The earlier papers close the field equations with prescribed kinematics. The magnetized models use a
linearly varying~\citep{Sahoo2017MSQM} or bilinear~\citep{Sahoo2018NewAst} deceleration parameter together with
$A\propto B^n$, which is inconsistent with a magnetized source (Sect.~\ref{sec:closure}). The models without a field use
a constant (Berman-law~\citep{Berman1983}) deceleration parameter together with $B=A^n$~\citep{AgrawalPawar2017}, a
parametrized deceleration parameter~\citep{Nagpal2018} or a power-law relation between $a$ and $q$~\citep{Pawar2025ObsQSQM}.
Their $q_0$ and $z_t$ are therefore inputs or functions of free constants. In Ref.~\citep{Sahoo2017MSQM}, for example,
$q_0=-0.73$ or $-0.65$ is taken from kinematic analyses, and $z_t$ ranges from $0.29$ to $0.82$ depending on a free constant
and on the adopted $q_0$. Here the closure is physical, and $q_0$ and $z_t$ are derived from fitted parameters. In GR, the MIT equation of
state has closed an LRS Bianchi~III model without a field, whose bag constant also acts as an effective vacuum
energy~\citep{Patra2026SQM}; together with flux freezing and the $\mu T$ coupling, the physical closure is new for MSQM in
$f(R,T)$. The second
difference concerns the fate and the anisotropy. The model of Ref.~\citep{Sahoo2017MSQM} ends in a Big Rip, and the
$A\propto B^n$ models keep a constant anisotropy~\citep{Sahoo2017MSQM,Sahoo2018NewAst}; the present model isotropizes and
approaches de~Sitter space. The third concerns the magnetic field, which drops out of the final solutions of the FLRW model
with $\Lambda$ of Ref.~\citep{AktasAygun2017}; here it is retained and obeys flux freezing. The fourth concerns the data.
Among the quark-matter $f(R,T)$ papers we examined, only Ref.~\citep{Pawar2025ObsQSQM} fits data, 57 $H(z)$ points for a
plane-symmetric model without a field, and its $H_0$ is consistent with ours (Table~\ref{tab:compare}). Within that set, the present work is the only one that confronts an MSQM model with a joint CC, Pantheon+ and DESI DR2 likelihood
with full covariances and compares it with $\Lambda$CDM through information criteria. A fit of the linear $f(R,T)$ model
to a perfect-fluid FLRW universe without quark matter or field, using the same classes of data, reaches a similar
verdict, a small AIC gain and a BIC penalty. For the model without bulk viscosity of Ref.~\citep{Saddiki2026DESI}, the
corrected AIC gives $-2.13$ and the BIC $+3.32$ relative to $\Lambda$CDM. Its coupling is not directly comparable with $\mu$,
because the matter content differs.

\begin{table}[ht]\centering\footnotesize
\setlength{\tabcolsep}{3pt}
\caption{Comparison of this work with flat $\Lambda$CDM and with earlier quark-matter cosmologies in $f(R,T)$ gravity.
``---'': not reported. Entries for earlier papers are as reported there; ``input'' marks values imposed rather than
derived. For Planck the values of $q_0$ and $z_t$ are derived from $\Omega_m=0.3153$ through $q_0=\tfrac32\Omega_m-1$ and
$1+z_t=(2\Omega_\Lambda/\Omega_m)^{1/3}$. The SH0ES $q_0$ is the value obtained in their Hubble-diagram fit
including SNe Ia at higher redshift~\citep{Riess2022}. The unit of $H_0$ is \kmsMpc. The uncertainties of this work are
symmetrized 68\% intervals (asymmetric values in Tables~\ref{tab:fit} and \ref{tab:derived}).}
\label{tab:compare}
\begin{tabular}{@{}>{\raggedright\arraybackslash}p{1.9cm}>{\raggedright\arraybackslash}p{2.3cm}>{\raggedright\arraybackslash}p{1.75cm}>{\raggedright\arraybackslash}p{1.5cm}>{\raggedright\arraybackslash}p{1.65cm}>{\raggedright\arraybackslash}p{1.2cm}>{\raggedright\arraybackslash}p{1.4cm}@{}}\toprule
Work & Model & Method & $H_0$ & $q_0$ & $z_t$ & Data\\\midrule
This work & MSQM, $f(R,T)=R+2\mu T$, LRS Bianchi~I (Kantowski--Sachs and Bianchi~III variants), MIT + Maxwell & field equations + MCMC & $69.1\pm3.2$ & $-0.496\pm0.027$ & $0.648\pm0.022$ & CC (32), Pantheon+ (1590), DESI DR2 (13)\\
Flat $\Lambda$CDM (this pipeline) & $\Lambda$CDM & MCMC & $69.9\pm3.2$ & $-0.544\pm0.012$ & $0.660\pm0.021$ & same as above\\
Flat $\Lambda$CDM, Planck 2018~\citep{Planck2018VI} & $\Lambda$CDM & CMB likelihood & $67.4\pm0.5$ & $-0.527$ (derived) & $0.632$ (derived) & Planck TT, TE, EE + lowE + lensing\\
SH0ES~\citep{Riess2022} & cosmography & distance ladder & $73.04\pm1.04$ & $-0.51\pm0.024$ & --- & Cepheids + SNe Ia\\
Sahoo et al.\ 2017~\citep{Sahoo2017MSQM} & MSQM, LRS Bianchi~I & linear $q(t)$, $A\propto B^n$ & --- & $-0.73$, $-0.65$ (input) & $0.29$--$0.82$ (free constant) & none\\
Sahoo et al.\ 2018~\citep{Sahoo2018NewAst} & MSQM with $\Lambda$, LRS Bianchi~I & bilinear or special $q(t)$, $A\propto B^n$ & --- & --- & --- (transition in $t$) & none\\
Akta\c{s} \& Ayg\"un 2017~\citep{AktasAygun2017} & MSQM, FLRW with $\Lambda$; field vanishes in the solutions & analytic solutions & --- & --- & --- & none\\
Agrawal \& Pawar 2017~\citep{AgrawalPawar2017} & quark and strange quark matter without field, plane symmetric & constant $q$ (Berman law), $B=A^n$ & --- & constant (input) & --- & none\\
Nagpal et al.\ 2018~\citep{Nagpal2018} & quark and strange quark matter without field, FLRW & parametrized $q$ & --- & --- & --- & none\\
Pawar et al.\ 2025~\citep{Pawar2025ObsQSQM} & quark and strange quark matter without field, plane symmetric & power-law $a$--$q$ relation; fit to $H(z)$ & $64.39^{+0.04}_{-0.47}$ & --- & --- & 57 $H(z)$\\\bottomrule
\end{tabular}
\end{table}

The sign of the fitted coupling depends on the choice of matter Lagrangian. For $L_m=-\rho$ the dust energy equation becomes
$(8\pi+\mu)\dot\rho_d=-(8\pi+2\mu)\theta\rho_d$, so $w_{d,\rm eff}=+\mu/(8\pi+\mu)$ instead of the $-\mu/(8\pi+3\mu)$ of
Eq.~\eqref{eq:rhod}. Distances at $z<2.4$ depend on $\mu$ only through $w_{d,\rm eff}$ (Sect.~\ref{sec:distances}), so the data
constrain $w_{d,\rm eff}$ independently of the Lagrangian, and for $L_m=-\rho$ the same posterior corresponds to
$\mu=-0.95^{+0.46}_{-0.43}$ (95\% interval $(-1.78,\,-0.03)$). The sign of $\mu$ is therefore tied to the Lagrangian
ambiguity~\citep{Fisher2019,HarkoMoraes2020,Carvalho2021}, whereas the physical result, the slower dust dilution of Sect.~\ref{sec:fitresults}, is not.

These results are subject to the following limitations. The coupling is assumed constant. If it were constant at all
epochs, matter--radiation equality ($\rho_d=\rho_r$) would move to $z_{\rm eq}=1494$ (68\%: 1030--2258), about half of
the Planck value $z_{\rm eq}=3402\pm26$~\citep{Planck2018VI}, because the slowly diluting dust [Eq.~\eqref{eq:rhod}] is less abundant
in the past than standard dust. The Planck value lies at the upper end of the 95\% interval (748--3535), which is reached only
as $\mu\to0$. A constant coupling of order unity is therefore in tension with the CMB unless it is confined to late times or
the early Universe is modified as well, and for the same reason no CMB likelihood is used and $r_d$ is free
(Sect.~\ref{sec:data}). The preference for $\mu>0$ depends on the SN sample, because it originates in the difference between
the SN and BAO matter densities, but not on the CC covariance (Table~\ref{tab:robust}). Adding the stellar-library term of
Ref.~\citep{Moresco2020}, or restricting the systematics to the 15 measurements for which the recipe was derived, changes
the median $\mu$ by at most 0.006 and the information criteria by less than 0.04. It affects
only the absolute scale ($H_0$, $r_d$, $t_0$); the medians of $\mu$ in Table~\ref{tab:robust} differ in the second decimal
only through rounding. Replacing
Pantheon+ by Union3~\citep{Rubin2025Union3} or DES-SN5YR~\citep{DES2024SN} raises the median $\mu$ and, with DES-SN5YR, gives a
Bayes factor in the weak range of the Jeffreys scale, but the recalibrated DES-Dovekie sample~\citep{Popovic2025Dovekie} of the same supernovae
returns values close to those of Pantheon+. The stronger preference is therefore not robust to the calibration, which
parallels the dependence of the DESI DR2 dark-energy preference on the SN sample (2.8$\sigma$, 3.8$\sigma$ and 4.2$\sigma$ with
Pantheon+, Union3 and DES-SN5YR)~\citep{DESIDR2}. The recalibration has the same effect on the dark-energy result: with CMB
and DESI DR2 BAO data, DES-Dovekie lowers the preference for evolving dark energy from the DES-SN5YR value to
$3.2\sigma$~\citep{Popovic2025Dovekie}, as it weakens our preference for $\mu>0$ (Table~\ref{tab:robust}). The Savage--Dickey Bayes factor is, moreover, proportional to the prior width
of $\mu$: it ranges from 0.65 for $\mathcal U(-3,3)$ to 2.2 for $\mathcal U(-10,10)$, and it is inconclusive for
all three priors. The model does not
explain why the late-time bag constant of Eq.~\eqref{eq:Bcfit} has the value of the dark-energy density. The field strength is not measured but is the CMB-motivated input
$\Bf_0=1$~nG, and the CMB shear limit used for comparison was derived for Bianchi~VII$_h$ templates. Finally, perturbations,
and hence structure formation and full stability, are not studied: a stability analysis requires the perturbation equations
of $f(R,T)$ gravity~\citep{Alvarenga2013}, and the sound-speed statements of Sect.~\ref{sec:EC} hold at the background level
only. The late-time result should therefore be read as a constraint on an effective late-time coupling, not as evidence for
a universal one.

\section{Conclusions}\label{sec:conclusions}
We have constructed and tested an MSQM cosmology in $f(R,T)=R+2\mu T$ gravity on the hypersurface-homogeneous background
of Eq.~\eqref{eq:metric}, with the field equations closed by the source itself rather than by a kinematic ansatz. The main
findings are the following.
\begin{enumerate}
\item \emph{Minimal Maxwell coupling.} The field enters Eq.~\eqref{eq:FE} with its general-relativistic weight, so the
$\mu T$ term modifies only the quark fluid and the dust, and the bag constant acts as the effective cosmological constant
$\Lambda_{\rm eff}$ (Sects.~\ref{sec:Lm} and \ref{sec:fieldeqs}).
\item \emph{Physical closure.} Kinematic closures are inconsistent with a magnetized source [Eq.~\eqref{eq:hA}], whereas the
bag-model equation of state and flux freezing close the system without any free function of time and admit the exact
solutions of Eqs.~\eqref{eq:isoexact} and \eqref{eq:kasnerdS} (Sects.~\ref{sec:closure} and \ref{sec:exact}).
\item \emph{Isotropization.} Without primordial shear, the field-induced anisotropy keeps a nearly constant radiation-era level
for $\mu\ge0$, as in GR, and then decays; the coupling changes only its dilution exponents, and all three curvature members
approach de~Sitter space (Sect.~\ref{sec:asymptotics}, Fig.~\ref{fig:shear}).
\item \emph{Observational constraints.} The joint CC+SN+BAO fit gives $\mu=1.12^{+0.65}_{-0.58}$. With $\Delta\mathrm{AIC}=-1.4$,
$\Delta\mathrm{BIC}=+4.0$ and an inconclusive Bayes factor, the data allow the coupling but do not require it
(Sect.~\ref{sec:fitresults}, Table~\ref{tab:comparison}).
\item \emph{Physical content.} For $\mu>0$ the dust dilutes more slowly than $a^{-3}$; the best-fit expansion is
quintessence-like and does not reproduce the evolution preferred by DESI DR2, and the NEC, WEC and DEC hold in every posterior
sample (Sects.~\ref{sec:res:DE} and \ref{sec:res:EC}).
\item \emph{Nature of the bag term.} The fitted bag term is an effective vacuum energy far below the QCD bag constant
[Eq.~\eqref{eq:Bcfit}], so strange quark matter does not cause the present acceleration (Sect.~\ref{sec:res:density}).
\end{enumerate}

These findings hold within the limitations of Sect.~\ref{sec:litcomp}. The coupling is an effective late-time
parameter, fitted without a CMB likelihood. Its posterior shifts with the SN sample, the Bayes factor scales with the prior,
the field strength is an input, and the matter-Lagrangian choice fixes the sign of $\mu$. Natural next steps are a
redshift-dependent coupling tested with CMB data, the perturbation theory of the model, and a quantitative treatment of the
QCD epoch with the physical bag constant.

\backmatter

\section*{Statements and Declarations}

\bmhead{Funding}
This research did not receive any specific grant from funding agencies in the public, commercial, or not-for-profit
sectors.

\bmhead{Competing interests}
The authors declare that they have no known competing financial interests or personal relationships that could have
appeared to influence the work reported in this paper.

\bmhead{Data availability}
This work uses publicly available data only: the 32 CC $H(z)$ measurements of the compilation of
Ref.~\citep{Moresco2022}, with the covariance of Ref.~\citep{Moresco2020}, the Pantheon+ SNe Ia
compilation~\citep{Scolnic2022,Brout2022}, the DESI DR2 BAO measurements~\citep{DESIDR2} and, for the robustness tests,
the DES-SN5YR~\citep{DES2024SN}, Union3~\citep{Rubin2025Union3} and DES-Dovekie~\citep{Popovic2025Dovekie} supernova
samples, all drawn from the cited references. No new observational data were generated. Sources, commit identifiers and
download dates are given in Appendix~\ref{app:mcmc}.

\bmhead{Code availability}
The Python code (symbolic derivations, solver, likelihoods, MCMC drivers and figure scripts), the Markov chains and the
processed data used in this work will be made publicly available in a repository with a Zenodo DOI upon publication; they
are available to the editor and referees on request.

\bmhead{Use of large language models}
A large language model was used to assist with language editing, drafting, symbolic checks and analysis code; all results
were verified by the authors, who take full responsibility for the content.

\bmhead{Author contributions}
D.\,P.~Tadas: Conceptualization. S.\,V.~Gore: Methodology, Software, Writing -- original draft, Writing -- review \&
editing. S.\,P.~Hatkar: Formal analysis. S.\,D.~Katore: Supervision.

\begin{appendices}
\section{Data, likelihood and sampling details}\label{app:mcmc}
\paragraph{Data provenance.} All data are public and were used unmodified. The files of the main analysis were downloaded on 17 October
2026 and the alternative SN samples on 8 November 2026; their SHA-256 checksums will be listed in the reproducibility package.
\begin{itemize}
\item \emph{Pantheon+}~\citep{Scolnic2022,Brout2022}: the files \texttt{Pantheon+SH0ES.dat} and
\texttt{Pantheon+SH0ES\_STAT+SYS.cov} from \url{https://github.com/PantheonPlusSH0ES/DataRelease} (commit
\texttt{c447f0f}). We keep the Hubble-flow sample of Sect.~\ref{sec:data} and use
$D_L=(1+z_{\rm hel})D_M(z_{\rm HD})$.
\item \emph{DESI DR2 BAO}~\citep{DESIDR2}: the Gaussian likelihood files (mean vector and covariance) distributed with
Cobaya, \url{https://github.com/CobayaSampler/bao_data}, directory \texttt{desi\_bao\_dr2} (commit \texttt{bb0c1c9}). These
give the 13 numbers listed in Sect.~\ref{sec:data}.
\item \emph{CC data}: the 32-point BC03 compilation of Ref.~\citep{Moresco2022}, with the original measurements of
Refs.~\citep{Jimenez2003,Simon2005,Stern2010,Moresco2012,Zhang2014,Moresco2015,Moresco2016,Ratsimbazafy2017,Borghi2022}.
The table and the systematic-error file of Ref.~\citep{Moresco2020} (SPS term: column \texttt{mod\_ooo}) were taken from a public mirror of M.~Moresco's
covariance repository, \url{https://github.com/Ahmadmehrabi/Cosmic_chronometer_data} (commit \texttt{350aad5}). The table
was cross-checked line by line against an independent compilation of the same data.
\item \emph{Alternative SN samples} (Table~\ref{tab:robust} only): DES-SN5YR~\citep{DES2024SN} (1829 SNe Ia, covariance
$C_{\rm sys}+{\rm diag}(\sigma^2_{\mu})$), Union3/UNITY1.5~\citep{Rubin2025Union3} (22 binned distance moduli) and
DES-Dovekie~\citep{Popovic2025Dovekie} (1820 SNe Ia, inverse covariance), all from the Cobaya SN data repository,
\url{https://github.com/CobayaSampler/sn_data} (commit \texttt{61d9643}), read as in the corresponding Cobaya likelihoods.
SN-only flat-$\Lambda$CDM fits reproduce the published $\Omega_m=0.352$, $0.356$ and $0.330$.
\end{itemize}

\paragraph{Likelihood.} The total likelihood is $-2\ln\mathcal L=\chi^2_{\rm CC}+\chi^2_{\rm SN}+\chi^2_{\rm BAO}$ of
Sect.~\ref{sec:data}, with the CC systematic terms of Eq.~\eqref{eq:CCcov} scaled by $H_{\rm fid}(z)$. The integral
$\int_0^z{\rm d}z'/E(z')$ is computed once per likelihood call by the trapezoidal rule with the end
correction $-\tfrac{1}{12}(\delta z)^2[f'(z)-f'(0)]$, $f=1/E$, on 6001 nodes in $0\le z\le2.6$. It is then evaluated at the
data redshifts by cubic Hermite interpolation, using the exact derivative $1/E$ at the nodes. The maximum relative error
against adaptive quadrature is $8.9\times10^{-11}$. The analytic marginalization over $M$ in Eq.~\eqref{eq:chi2SN} agrees
with numerical profiling to $10^{-6}$.

\paragraph{Sampling.} The sampler was \textsc{emcee}~3.1.6~\citep{ForemanMackey2013}
with 48 walkers and $1.2\times10^4$ steps per run, using differential-evolution moves (weight 0.8) and snooker moves (weight 0.2)
\citep{terBraak2006,terBraakVrugt2008}. Walkers start from a Gaussian distribution centered on
$(H_0,\Omega_{d0},\mu,r_d)=(68,\,0.30,\,0,\,147)$ with widths $(2,\,0.02,\,0.3,\,3)$, with $H_0$ in \kmsMpc{} and $r_d$ in Mpc. The eight runs (two models, four
data combinations) use the fixed seeds 11, 12, 21, 22, 31, 32, 41 and 42; the robustness runs of Table~\ref{tab:robust} use
the same settings and the seeds 11 (MSQM) and 12 ($\Lambda$CDM). Convergence was monitored with the integrated
autocorrelation time $\tau_{\rm int}$: every run satisfies $N_{\rm steps}>50\,\tau_{\rm int}^{\max}$, where
$\tau_{\rm int}^{\max}$ is the largest integrated autocorrelation time; the first $5\tau_{\rm int}^{\max}$ steps are discarded as
burn-in, and the chains are thinned by $\tau_{\rm int}^{\max}/2$. The split Gelman--Rubin statistic
$\hat R$~\citep{GelmanRubin1992,Vehtari2021} is computed from 96 half-walker chains. Because the walkers of an ensemble sampler
are not independent chains, it serves only as a heuristic; convergence is assessed primarily by the run length, which exceeds
100 integrated autocorrelation times for every parameter in every run. All eight runs have converged
(Table~\ref{tab:convergence}): every run has $N_{\rm steps}/\tau_{\rm int}^{\max}>330$ and
$\hat R-1<0.005$, well inside the usual thresholds $N_{\rm steps}/\tau_{\rm int}>50$ and $\hat R-1<0.01$. For the
joint fit, $\tau_{\rm int}^{\max}=34$ steps, $N_{\rm steps}/\tau_{\rm int}^{\max}=358$ and
$\hat R-1<0.003$ for every parameter. Best fits are found by repeated Nelder--Mead minimization of $\chi^2$,
started at the sample of highest posterior. Derived quantities come from 1500 randomly drawn posterior samples, each
propagated through the full anisotropic solver (DOP853, $r_{\rm tol}=10^{-10}$, from $a=10^{-20}$).

\begin{table}[ht]\centering\footnotesize
\setlength{\tabcolsep}{3pt}
\caption{Robustness of the joint CC+SN+BAO fit of the MSQM model. Upper block: treatments of the CC systematics with
Pantheon+; (f) is the baseline (Sect.~\ref{sec:data}). Lower block: alternative SN samples, with the CC covariance (a),
which differs from (f) only in the scaling of the systematic terms. The unit of $H_0$ is \kmsMpc; medians with 68\%
intervals; $\Delta$ values are MSQM minus flat $\Lambda$CDM fitted to the same data; $B_{01}$ is the Savage--Dickey Bayes
factor for $\mu=0$ with the prior $\mathcal U(-6,6)$.}
\label{tab:robust}
\begin{tabular}{@{}>{\raggedright\arraybackslash}p{3.3cm}cccrrrc@{}}\toprule
Variant & $H_0$ & $\mu$ & $P(\mu>0)$ & $\Delta\chi^2$ & $\Delta$AIC & $\Delta$BIC & $B_{01}$\\\midrule
(f) IMF+SPS, fiducial scaling (baseline) & $69.1\pm3.2$ & $1.12^{+0.65}_{-0.58}$ & $0.98$ & $-3.4$ & $-1.4$ & $+4.0$ & $1.30$\\
(a) IMF+SPS, observed $H$ scaling & $67.9\pm3.2$ & $1.12^{+0.65}_{-0.58}$ & $0.98$ & $-3.4$ & $-1.4$ & $+4.0$ & $1.23$\\
(b) IMF+st.\,lib.+SPS & $64.2\pm5.2$ & $1.12^{+0.68}_{-0.58}$ & $0.98$ & $-3.4$ & $-1.4$ & $+4.0$ & $1.21$\\
(g) as (b), fiducial scaling & $69.1\pm5.4$ & $1.12^{+0.66}_{-0.59}$ & $0.98$ & $-3.4$ & $-1.4$ & $+4.0$ & $1.30$\\
(c) systematics on 15 points only & $68.9\pm2.4$ & $1.13^{+0.65}_{-0.58}$ & $0.98$ & $-3.4$ & $-1.4$ & $+4.0$ & $1.25$\\\midrule
Union3 instead of Pantheon+ & $67.4\pm3.2$ & $1.49^{+0.84}_{-0.72}$ & $0.99$ & $-4.0$ & $-2.0$ & $+0.2$ & $0.74$\\
DES-SN5YR instead of Pantheon+ & $67.2\pm3.2$ & $1.66^{+0.68}_{-0.62}$ & $0.998$ & $-7.7$ & $-5.7$ & $-0.2$ & $0.15$\\
DES-Dovekie instead of Pantheon+ & $67.8\pm3.3$ & $1.15^{+0.62}_{-0.56}$ & $0.99$ & $-4.0$ & $-2.0$ & $+3.5$ & $0.92$\\
\bottomrule
\end{tabular}
\end{table}

\begin{table}[ht]\centering\small
\caption{Convergence of the MCMC runs (48 walkers, $1.2\times10^4$ steps each).}
\label{tab:convergence}
\begin{tabular}{@{}llccccc@{}}\toprule
Model & Data & $\tau_{\rm int}^{\max}$ & $N_{\rm steps}/\tau_{\rm int}^{\max}$ & burn-in & $\max(\hat R-1)$ & acceptance\\\midrule
MSQM & CC+SN+BAO & 33.5 & 358 & 168 & 0.0029 & 0.20\\
MSQM & CC+SN & 14.2 & 842 & 72 & 0.0011 & 0.25\\
MSQM & CC+BAO & 35.8 & 335 & 180 & 0.0042 & 0.20\\
MSQM & CC & 16.3 & 735 & 82 & 0.0016 & 0.23\\
$\Lambda$CDM & CC+SN+BAO & 25.1 & 478 & 126 & 0.0022 & 0.22\\
$\Lambda$CDM & CC+SN & 9.0 & 1341 & 45 & 0.0008 & 0.31\\
$\Lambda$CDM & CC+BAO & 25.2 & 476 & 126 & 0.0026 & 0.22\\
$\Lambda$CDM & CC & 10.4 & 1158 & 52 & 0.0009 & 0.29\\
\bottomrule
\end{tabular}
\end{table}

\paragraph{Software.} Python with NumPy, SciPy, SymPy, Matplotlib and GetDist; exact versions will be listed in the reproducibility package.

\end{appendices}

\bibliography{references}

\end{document}